\documentclass[a4paper,amsmath,superscriptaddress,twocolumn,showkeys,showpacs,prb,aps, 10pt]{revtex4-1}
\usepackage{color}
\usepackage{graphicx}
\usepackage{hyperref}

\newcommand{\dagga}{{\phantom{\dagger}}}

\newcommand{\be}{\begin{equation}}
\newcommand{\ee}{\end{equation}}
\newcommand{\bea}{\begin{eqnarray}}
\newcommand{\eea}{\end{eqnarray}}
\newcommand{\ba}{\begin{eqnarray*}}
\newcommand{\ea}{\end{eqnarray*}}
\newcommand{\up}{\uparrow}
\newcommand{\down}{\downarrow}
\newcommand{\eqn}[1]{(\ref{#1})}
\newcommand{\fract}[2]{\frac{\displaystyle #1}{\displaystyle #2}}

\begin{document}

\title{From density functional to Kondo: magnetic impurities in nanotubes }

\author{P. P. Baruselli}
\affiliation{SISSA, Via Bonomea 265, Trieste 34136, Italy}
\affiliation{CNR-IOM, Democritos Unit\'a di Trieste, Via Bonomea 265, Trieste 34136, Italy}
\affiliation{Institut f\"{u}r Theoretische Physik, Technische Universit\"{a}t Dresden, 01062 Dresden, Germany}

\author{M. Fabrizio}
\affiliation{SISSA, Via Bonomea 265, Trieste 34136, Italy}
\affiliation{CNR-IOM, Democritos Unit\'a di Trieste, Via Bonomea 265, Trieste 34136, Italy}

\author{A. Smogunov}
\affiliation{Voronezh State University, University Square 1, Voronezh 394006, Russia}
\affiliation{CEA, IRAMIS, SPCSI, F-91191 Gif-sur-Yvette Cedex, France}

\author{R. Requist}
\affiliation{SISSA, Via Bonomea 265, Trieste 34136, Italy}

\author{E. Tosatti}
\affiliation{SISSA, Via Bonomea 265, Trieste 34136, Italy}
\affiliation{CNR-IOM, Democritos Unit\'a di Trieste, Via Bonomea 265, Trieste 34136, Italy}
\affiliation{ICTP, Strada Costiera 11, Trieste 34014, Italy}

\date{\today}

\begin{abstract}
Low temperature electronic conductance in nanocontacts, scanning tunneling microscopy (STM),
and metal break junctions involving 
magnetic atoms or molecules 
is a growing area with important unsolved theoretical problems.  
While the detailed relationship between contact geometry and electronic structure
requires a quantitative ab initio approach such as density functional theory (DFT), the 
Kondo many body effects ensuing from the coupling of the impurity spin with metal electrons 
are most properly addressed by formulating a generalized Anderson impurity model 
to be solved with, for example, the numerical renormalization group (NRG) method.  Since there is
at present no seamless scheme that can accurately carry out that program, we have 
in recent years designed a systematic method for semiquantitatively joining DFT and NRG. 
We apply this DFT-NRG scheme to the ideal conductance of single wall (4,4) and (8,8) nanotubes with magnetic 
adatoms (Co and Fe), both 
inside and outside the nanotube, and with a single carbon atom vacancy. 
A rich scenario emerges, with Kondo temperatures generally in the Kelvin range, and
conductance anomalies ranging from a single channel
maximum to destructive Fano interference with cancellation of two channels out of the total 
four.  The configuration yielding the highest Kondo temperature (tens of Kelvins) and
a measurable zero bias anomaly is that of a Co or Fe impurity inside
the narrowest nanotube. The single atom vacancy has a spin, but a very low
Kondo temperature is predicted. The geometric, electronic, and symmetry factors influencing this variability
are all accessible, which makes this approach methodologically instructive and highlights many
delicate and difficult points in the first principles modeling of the Kondo effect
in nanocontacts.  

\end{abstract}

\maketitle

\section{Introduction}

When the contact between two metals is reduced to the ultimate monoatomic limit---a geometry realized in 
mechanical break junctions \cite{agrait}, but also in STM \cite{Neel2007}---the 
electrical conductance can be satisfactorily understood and calculated by applying Landauer's
standard ballistic formalism~\cite{jauho_wm} 
to an equally standard ab initio electronic 
structure calculation of the nanocontact~\cite{larade};
for an alternative formulation, see Ref.~\onlinecite{tsukada}.
However, when a magnetic atom (such as Co) or magnetic molecule 
(such as Cu-phthalocyanine) bridges two nonmagnetic metallic leads, the conductance reflects the presence of the
impurity spin and its Kondo screening. The characteristic Kondo signature is a low voltage conductance 
peak, or dip, present with or without a magnetic field and generally referred to as a zero bias anomaly
~\cite{appelbaum_pr, andersontunnel,gupta}. 

The zero bias anomaly is determined by the electronic
structure of the nanocontact.  
Given the atomic nature of a nanocontact, as opposed to the smoothness 
of mesoscopic contacts such as quantum dots \cite{kastner}, a
quantitative ab initio approach is mandatory to represent the 
geometry-dependent electronic structure, the local spin density, etc., in realistic detail. 
That information is available, albeit approximately, from spin-polarized density functional theory (DFT) calculations but comes at the price of breaking spin-rotational symmetry.   
Spontaneous 
spin-rotational symmetry breaking does indeed occur in infinite
magnetic systems,
which DFT describes reasonably well,  
but not in a single magnetic atom, molecule, or dot. As a result, spin-polarized DFT completely misses 
the Kondo screening of the local magnetic moment by the leads 
~\cite{Kondo1964}, 
thus failing to provide the correct low temperature low field conductance 
and zero bias anomaly. A full description of the 
Kondo physics requires instead an explicit many-body technique, such as NRG \cite{wilson}. 
Although promising approximate ab initio based approaches have been proposed \cite{thygesen, jacob}, 
a quantitatively accurate description of Kondo physics has only been achieved with NRG.  But due to the 
complexity of the problem, NRG-type methods cannot handle all the electronic degrees of freedom of a 
realistic lead-impurity-lead contact geometry and are only practical for highly simplified Anderson impurity 
models (AIM) \cite{anderson61,hewson}, whose parameters could only be estimated phenomenologically thus far, 
leaving us without a quantitative ab initio based 
method for the prediction of magnetic nanocontact conductance, 
even at zero temperature, 
low voltage and zero field. 
To be sure, several important discussions are present in the literature where DFT electronic structure 
calculations have been employed to argue qualitatively for a given impurity spin, and/or where NRG calculations
have been used to distinguish the different temperature and field behavior predicted
for different spins. ~\cite{wehling,Costi2009,potok_overscreened, Costi-magnetic-field-underscreened,Parks11062010,rohlfing} 
What is however still needed
is an approach where geometrical and orbital complications are included at the outset and 
connected to subsequent NRG calculations at a quantitative level.

Here we present an implementation, based 
on work recently developed in our group~\cite{lucignano},  which attempts to 
improve this situation by means of a well defined semiquantitative scheme for joining DFT and NRG. 

The scheme is straightforward. 
The basic consideration is that a spin-polarized DFT calculation of a magnetic impurity can be regarded as 
conceptually similar to the mean-field treatment of a generalized AIM.  Like the Hartree-Fock solution of the original AIM \cite{anderson61}, it provides a mean-field rationale for the existence of free local moments in transition metal impurities and alloys 
in a nonmagnetic host metal. 
Furthermore, AIMs neglect the interactions between conduction electrons, which parallels the underlying assumption of the local density approximation (LDA) and generalized gradient 
approximation (GGA) that such interactions only 
modify the band structure parameters, i.e.~that the host metal can be described by 
noninteracting quasiparticles with an effective band dispersion.
On the basis of this correspondence, we assume there exists an AIM that reproduces, within mean field, the ab initio results for a generic nanocontact. 
The key point is to select which particular ab initio 
quantities 
the AIM mean field should reproduce in order for the AIM itself
to provide the best possible description of the low temperature conductance through the nanocontact.  
Concerning the internal 
electronic degrees of freedom of the impurity, the choice is 
practically mandatory and is dictated by the localized orbitals that are primarily involved 
in magnetism. For instance, in the case of a transition 
metal atom one must assume that the model 
includes at least the $d$ orbitals (and potentially one $s$ orbital).  
The choice of conduction channels 
among the electronic states 
of the free leads 
is similarly mandatory and dictated
by the requirement that they 
should share 
the same symmetry as the impurity 
states with which they hybridize. 

What is not equally straightforward is to find 
unambiguous 
criteria determining the impurity-lead 
hybridization parameters and the interaction terms acting within the impurity. One problem is that 
the ab initio results are obtained by explicitly breaking spin $SU(2)$ symmetry, whereas the 
parameters we are seeking belong to a spin-rotationally invariant Hamiltonian. 
For instance, 
the spectral density of states of the magnetic impurity, determined using a basis of localized orbitals, has a strong spin-splitting and is generally peaked far away from the chemical potential in spin-polarized DFT, making it difficult to reconstruct the spin-rotationally invariant hybridization functions that enter the AIM. 
Conversely, the alternative possibility of starting from a spin unpolarized calculation would yield 
too little information about the spin state of the impurity.  For example, it would not tell us whether 
a d$^7$ impurity, say Co$^{2+}$, in an octahedral environment has a high spin or low spin state, 
namely $S=3/2$ or $S=1/2$.
%

Our approach toward fixing the model parameters is to make use of the additional information contained 
in the ab initio scattering phase shifts \cite{lucignano}.  Specifically, given a particular nanocontact geometry, 
one can identify the symmetry-adapted scattering eigenchannels and, for each of them, calculate the scattering 
phase shifts for any spin projection relative to the direction of the mean field magnetization.  We shall see in 
Sec.~\ref{sec_anderson} that these phase shifts together with a few other quantities from the ab initio 
calculation will allow us to determine an effective AIM relatively unambiguously. The subsequent solution of 
this model by NRG provides results that we believe are representative of the low temperature behavior of the 
realistic nanocontact, including in particular the low bias conductance anomalies.  

The choice of single wall carbon nanotubes for our application was guided 
not so much by experiment, which is 
still
missing, but rather by the 
consideration that 
nanotubes 
possess well-defined one dimensional conduction
channels.  The system's intrinsic simplicity and robustness makes it an ideal test case
for a thorough study. 
Although a first exploratory preview was recently presented ~\cite{ baruselli_prl2012},
here we now provide a full accont of the protocol and of its results, including particularly 
carbon vacancies which here play the unusual role of magnetic impurities .

The paper is organized as follows. In section \ref{sec_dft} we present DFT results for Co and Fe 
impurities on (4,4) and (8,8) nanotubes. In section \ref{sec_anderson} we set up the AIM and explain
how to fix its parameters. 
Sections \ref{sec_results} and \ref{sec_fe} present our NRG results concerning Kondo behavior, respectively 
for Co and Fe impurities. In section \ref{sec_zba} we report some results about the zero bias anomalies 
to be expected in such systems. The additional case of a nanotube vacancy acting as a magnetic 
impurity  is presented in section \ref{sec_vac}, including both DFT and NRG results. Finally, in section 
\ref{sec_conclusions} we draw the main conclusions of this work.

\section{Cobalt and iron impurities on nanotubes: DFT study}\label{sec_dft}

\subsection{Electronic structure}
\begin{figure}[tbp]\centering
\includegraphics[width=0.4\textwidth,angle=0]{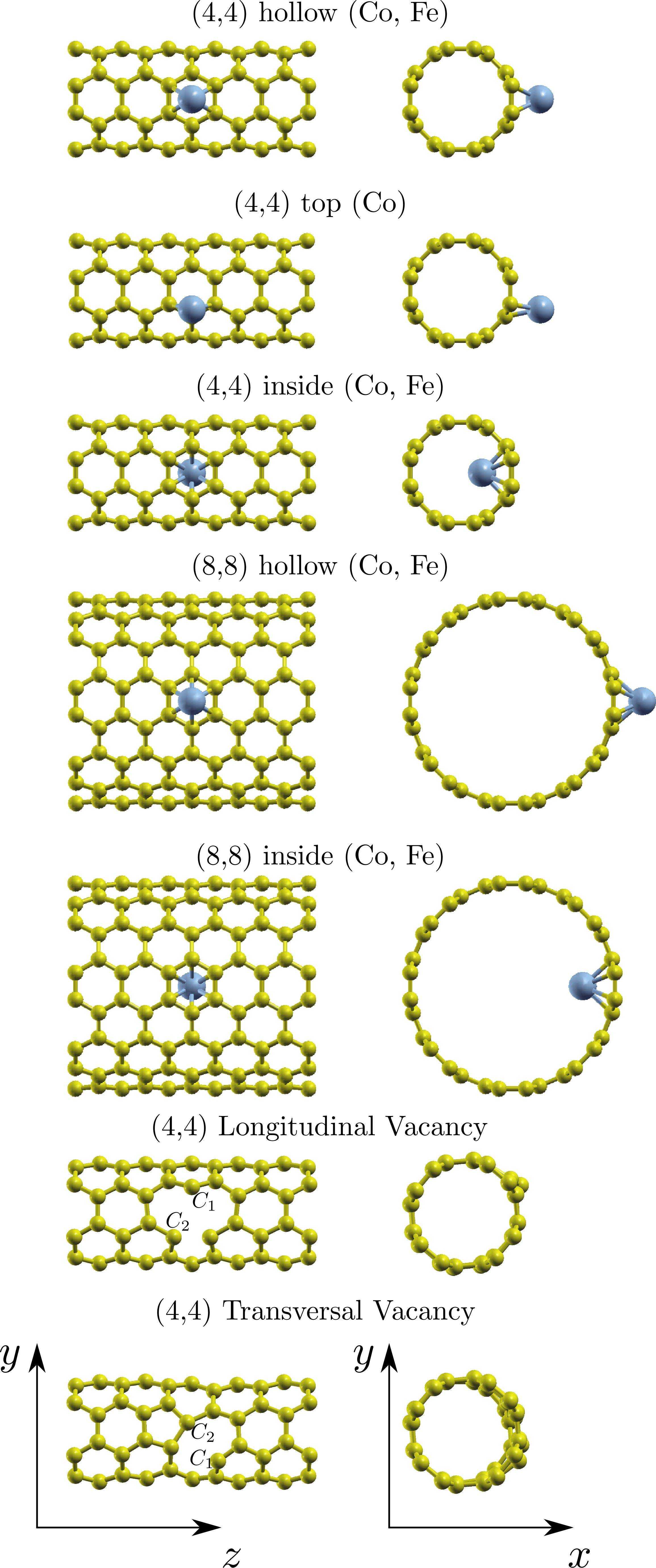}
\caption{
Co and Fe adatoms on single wall carbon nanotubes, and a single atom vacancy in two inequivalent relaxed configurations.
}
\label{fig_struc}
\end{figure}

Our study starts with electronic structure calculations, greatly extending previous ones, 
Refs.~\onlinecite{Baruselli_physE, baruselli_prl2012} 
for single Co or Fe atoms adsorbed on a metallic single-wall carbon nanotube (SWNT) (Fig.~\ref{fig_struc}). 
In order to study the effect of the nanotube curvature, we considered two different nanotubes, (4,4) and (8,8), of different radius.
We did not consider a Ni adatom since, as 
recently reported \cite{yagi}, it generally loses its magnetic moment when adsorbed on carbon nanotubes.

We begin by defining the scattering region, which we take to be a nanotube segment consisting of $N_C$  carbon atoms ($N_C =80, 160$ for 
(4,4) and (8,8) tubes, respectively) and one impurity.
For this system, we  first 
carried out 
standard DFT calculations with periodic boundary conditions and relaxed the positions 
of all the atoms in the unit cell shown in Fig.~\ref{fig_struc}, 
except those in the two outermost rings, to improve the convergence toward the infinite tube limit.
Calculations were performed with the plane wave package Quantum ESPRESSO \cite{QE-2009} 
within the GGA to the exchange-correlation energy in the parametrization of
Perdew, Burke and Ernzerhof~\cite{pbe}. The planewave cutoffs 
were 30 Ry and 300 Ry for the wave functions and the charge density, 
respectively. Integration over the one-dimensional Brillouin zone 
was accomplished using 
$8$ $k$-points and a smearing parameter of $10$ mRy. 
When necessary to test the effects of electron correlations and self-interaction errors, we extended the calculations 
from GGA to GGA+$U$, including a Hubbard $U$ interaction 
within 
the transition metal $d$ orbitals.
While this was occasionally important as a check on the stability of the impurity spin state, most results presented below 
were obtained in the GGA. 

Our DFT calculations 
suggest 
that in all cases the hollow site 
is the most stable adsorption configuration. For example, the ontop (external) Co adsorption 
configuration was about 47 meV higher in energy that the hollow site on the (4,4) SWNT. 
Nevertheless, 
we also included in our study 
the case of a Co adatom adsorbed at the ontop position
of a (4,4) SWNT 
to gain insight into the  
influence of 
adsorption site on the magnetic and transport properties of the nanotube. Also, 
although such adsorption geometries are higher in energy, they might still be accessible in experiment.
In order to 
explore 
the possible role of 
self-interaction errors, we performed GGA+$U$ calculations for the selected case of hollow-site Co on the (4,4) SWNT.
With a value of $U=2$ eV for $d$ orbitals of Co, 
we did not find meaningful changes of the $S=1/2$ state of the Co adatom.
%
Table \ref{table_struc} summarizes the results of the geometry relaxation
for all the systems studied and also reports the total spin magnetic moment for each case.
These results 
compare well with those reported recently 
by Yagi and co-workers \cite{yagi}.

\begin{figure}[tbp]\centering
\includegraphics[width=0.35 \textwidth,angle=0]{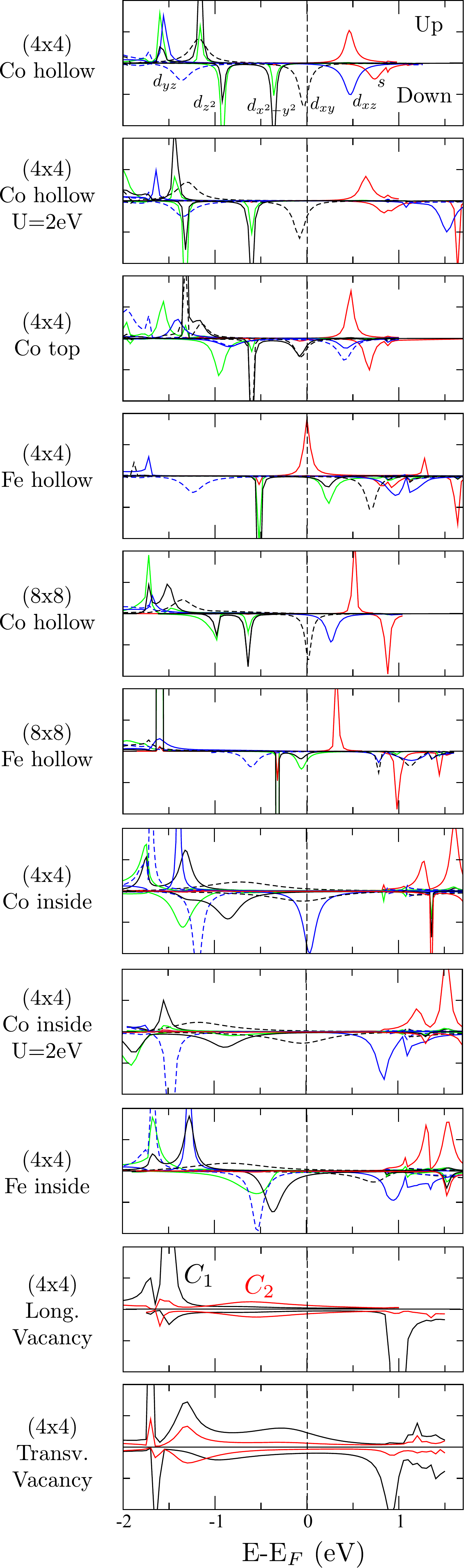}
\caption{
PDOS on the impurity atomic orbitals. The curves on the upper panel are marked by 
the corresponding atomic character. Dashed vertical lines indicate the position of the Fermi level. We do not show the case of Co and Fe inside the (8,8) SWNT, but it is qualitatively similar to the case inside the (4,4) SWNT.
In the case of vacancies, the total PDOS on $2s$ and $2p$ orbitals of atoms denominated $C_1$ (the lone atom) and $C_2$ (one of the other two atoms originally close to the removed atom - see Fig.~\ref{fig_struc}) is given.
}
\label{fig_pdos}
\end{figure}

Fig.~\ref{fig_pdos} presents the projected density of states (PDOS) for the $s$ and $d$ orbitals
of the TM adatom. 
The different curves, labeled in the upper panel, correspond to the character of the orbital. 
The PDOS 
shows sharp spin-split peaks corresponding to  
the magnetic orbitals of the TM atom that will subsequently be
used to construct the many-body model
Hamiltonian. 
The crucial element here is symmetry.
All the TM orbitals can be classified according to their symmetry as follows. 
In both hollow and ontop geometries, there is a mirror plane $xy$ 
through the TM atom and orthogonal to the nanotube (see Fig.~\ref{fig_struc}).
The states are therefore either even ($e$) or odd ($o$) with respect to the corresponding reflection operation.  
For the hollow adsorption site, there is in addition the symmetry plane $xz$. 
We can assign therefore an extra index $s$ (symmetric) or $a$ (antisymmetric) to 
states which are even or odd with respect to this additional symmetry plane. 
As an example, 
consider the Co atom adsorbed at the hollow site of the (4,4) nanotube (upper panel of Fig.~\ref{fig_pdos}). 
In this case there is only one magnetic orbital, $d_{xz}$,
which has the $\{o,s\}$ symmetry and is singly occupied by a spin up electron in our DFT calculations.
All other $d$ orbitals are fully occupied and therefore irrelevant for low temperature physics, including
the $d_{xy}$ orbital, which, partially empty in the spin down channel, becomes almost fully occupied
when a finite $U=2$ eV is introduced in the calculation (see the second panel from the top in Fig.~\ref{fig_pdos}).
This concludes the analysis of the relevant impurity orbitals and their symmetry.

\begin{figure}[tbp]\centering
\includegraphics[width=0.4\textwidth,angle=0]{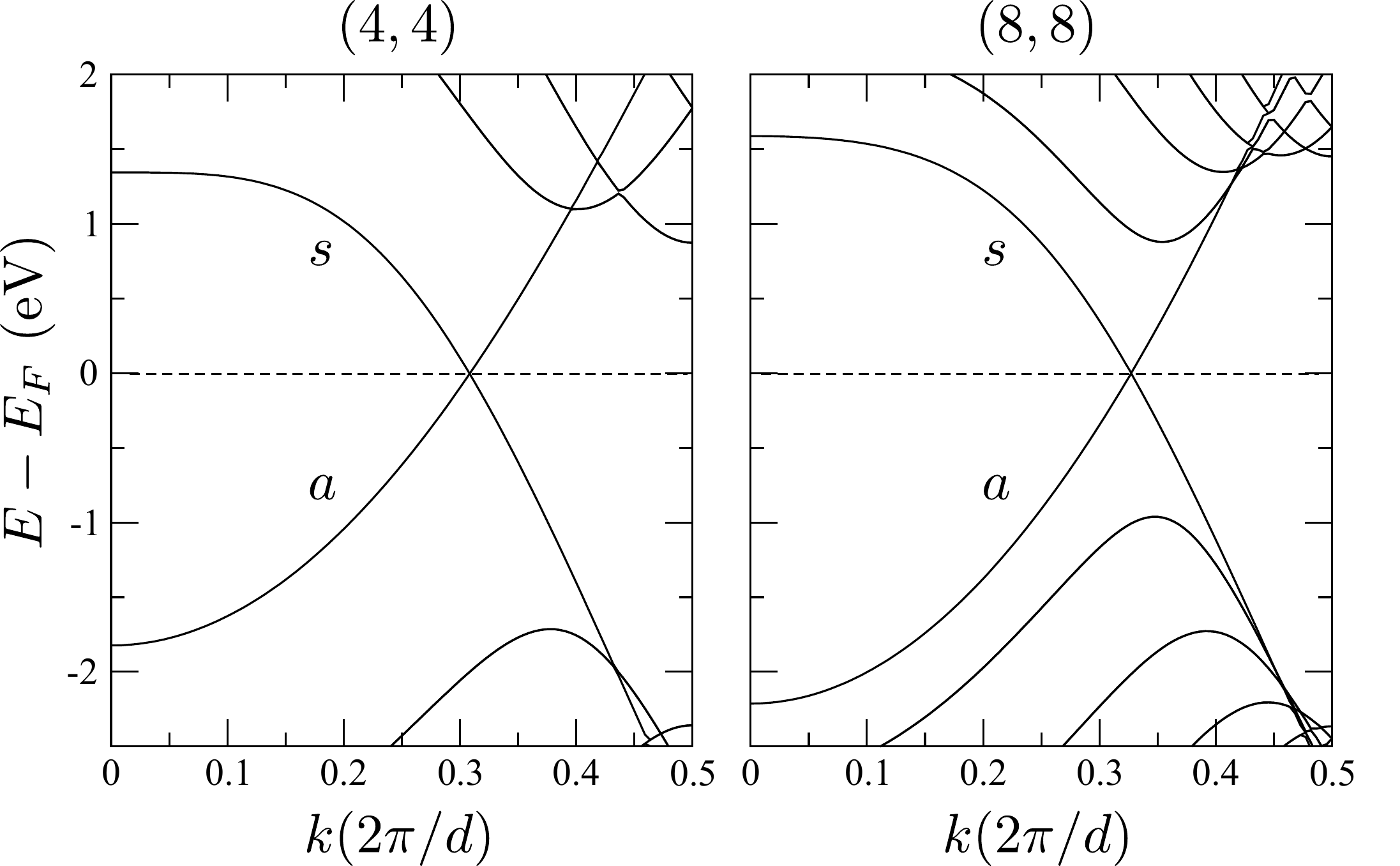}
\caption{
Electronic bands of pure (4,4) and (8,8) SWNTs. The two bands crossing the Fermi energy
are labeled as $s$ (symmetric) and $a$ (antisymmetric) which reflects their symmetry 
with respect to the mirror plane $xz$ (see Fig.~\ref{fig_struc}). 
}
\label{fig_bands}
\end{figure}

The next step is to identify the nanotube conduction channels carrying the electrons which will 
scatter on the magnetic impurity.  This is done by examining the electronic structure of the
infinite, impurity-free nanotube.
Figure~\ref{fig_bands}, shows the band structure of (4,4) and (8,8) carbon nanotubes with, in both cases, 
two conduction bands crossing the chemical potential.   One is symmetric and 
the other antisymmetric with respect to the mirror $xz$-plane. We label them 
as $s$ and $a$ in accordance with the 
above notation. 
Each of these two bands has left- and right-moving states, $\phi_l$ and $\phi_r$, 
which can be combined to form even and odd combinations, $\phi_{e/o}=(\phi_l\pm\phi_r)/\sqrt2$. 
The four resulting conduction channels, which can be labeled by the pair $\{e/o,s/a\}$, 
identify the four scattering channels that will couple to impurity orbitals of same symmetry.

\begin{table}\centering
\begin{tabular}{|l|c|c|}

\hline
Configuration   & TM-C dist. (\text{\AA}) & $\mu$ ($\mu_{B}$) \\
\hline
(4x4) Co hollow               & 2.07 (4), 2.32 (2)  & 1.26  \\
(4x4) Co hollow, $U = 2$ eV       & 2.08 (4), 2.33 (2)  & 1.17  \\  
(4x4) Co ontop ($\Delta E=47$ meV)  & 1.99 (2), 2.00 (1)  & 1.16  \\
(4x4) Co inside ($\Delta E=193$ meV)  & 2.15 (4), 1.94 (2)  & 0.79 \\
(4x4) Co inside, $U = 2$ eV  & 2.19 (4), 1.97 (2)  & 1.01 \\
(4x4) Fe hollow               & 2.17 (4), 2.42 (2)  & 3.40  \\ 
(4x4) Fe inside ($\Delta E= 199$ meV)  & 2.15 (4), 1.98 (2)  & 1.84 \\
(8x8) Co hollow               & 2.09 (4), 2.21 (2)  & 1.32  \\
(8x8) Fe hollow               & 2.11 (4), 2.23 (2)  & 2.35  \\
(8x8) Co inside ($\Delta E=53$ meV)    & 2.11 (4), 2.02 (2)  & 1.18  \\
(8x8) Fe inside ($\Delta E=130$ meV)    & 2.13 (4), 2.05 (2)  & 2.15  \\
(4x4) Long. vac. ($\Delta E=0.8$ eV)    & 1.37 (2), 2.86 (2)  & 1.05  \\
(4x4) Transversal vacancy     & 1.39 (1), 1.39 (1),   & 0.89  \\
&2.64 (1), 2.70 (1)&\\

\hline
\end{tabular}
\caption{
Geometry and spin magnetic moment of Co and Fe
adatoms on different SWNTs. The number of equivalent C atoms with the same nearest-neighbor
distance to the adatom is given in parentheses; in the case of vacancies, the distance of the lone $C_1$ atom to its nearest neighbors is given. The last column gives
the total spin magnetic moment obtained by integrating the spin magnetization over
the whole unit cell. 
}\label{table_struc}
\end{table}

\begin{figure}[tbp]\centering
\includegraphics[width=0.4\textwidth,angle=0]{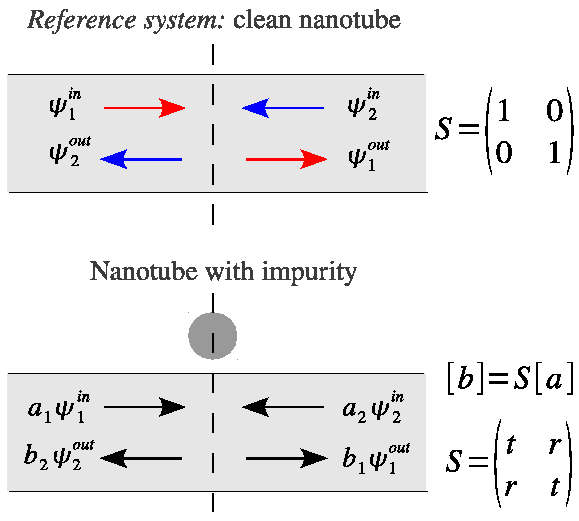}
\caption{
Schematic picture explaining the calculation of even and odd phase shifts $\delta_{e/o}$.
They appear after diagonalizing of the scattering matrix $S$. 
The clean nanotube without impurity is the reference system defining the incoming 
and outgoing waves with respect to which the scattering matrix
is calculated when the impurity is introduced.    
}
\label{fig_sketch}
\end{figure}

\begin{figure}[tbp]\centering
\includegraphics[width=0.35\textwidth,angle=0]{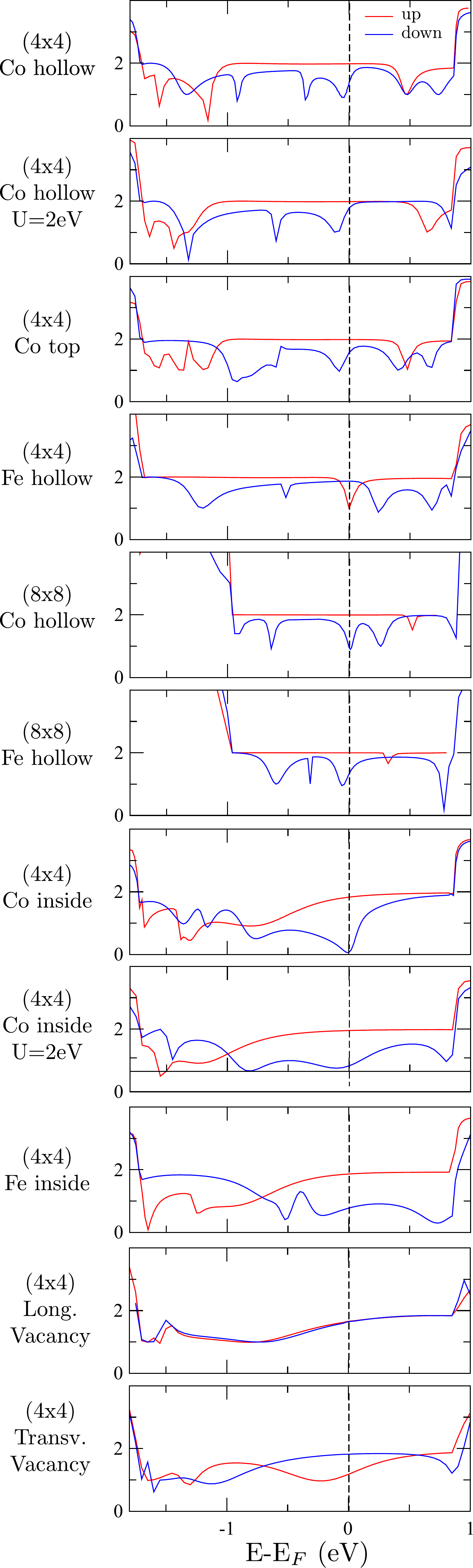}
\caption{
Spin-polarized transmission functions for the same cases as those in Fig. \ref{fig_pdos}.
 }
\label{fig_t}
\end{figure}

\begin{figure}[tbp]\centering
\includegraphics[width=0.4\textwidth,angle=0]{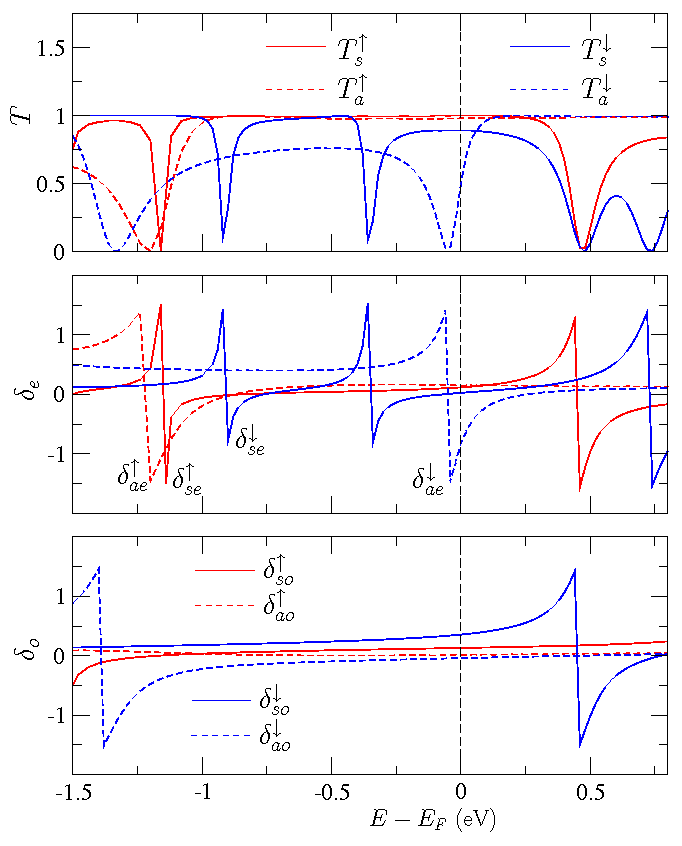}
\caption{
Spin-polarized transmission eigenvalues (upper panel) and even (middle panel) and odd (lower panel) 
phase shifts, $\delta_{e/o}$, for a Co adatom on (4,4) SWNT (hollow site). Solid and dashed
lines correspond to symmetric and antisymmetric channels (with respect to the $xz$ plane). 
The curves for spin up and spin down polarizations are shown in red and blue color. }
\label{fig_tdelta}
\end{figure}

\subsection{Transmission function and phase shifts from density functional calculations }

The main physical property of interest to us is the linear electrical conductance  
near zero bias of the nanotube with a single magnetic impurity. 
Within the mean-field 
DFT scheme,
which is the initial stage of our approach, 
the linear response ballistic conductance is given by the Landauer-Buttiker formula,
$G=(G_0/2) T(E_F)$, where $T(E_F)$ is the total electron transmission 
at the Fermi level and $G_0=2e^2/h$ is the conductance quantum. 
In our collinear spin-polarized calculations, the total transmission is just the sum of the two independent spin channels, $T(E_F)=\sum_{\sigma}T^{\sigma}(E_F)$.
The transmission function in each spin channel is given by the trace (suppressing the spin index and the energy argument),
$T={\rm tr}({\bf t}^{\dagger}{\bf t})$, where $t_{ij}$ is in our case the $(2\times2)$ matrix of transmission amplitudes
with $i,j=s$ or $a$. For the hollow adsorption site this matrix is diagonal since scattering conserves 
reflection
symmetry in the $xz$ plane, and therefore $T=|t_s|^2+|t_a|^2$, where $t_{s/a}$ are the transmission amplitudes 
for the two independent channels.
It should be stressed here that the mean-field transmission and the associated conductance
are simply an intermediate calculational step and by no means represent our final
conductance result, which will follow the NRG study.

In order to calculate the transmission amplitudes, we take the unit 
cell of Fig.~\ref{fig_struc} as the scattering region and smoothly attach
semi-infinite carbon nanotubes to both sides. 
The transmission and reflection amplitudes of such an open system are then calculated 
using a wave function matching approach~\cite{2004Smogunov-PRB} implemented in the  PWCOND code, 
which is part of the Quantum ESPRESSO package.

We present in Fig.~\ref{fig_t} spin-dependent transmission functions for all cases under consideration.
Around the Fermi energy the total transmission per spin 
has a maximum value of two, corresponding to the two available electron bands. 
As a function of energy,
the transmission curves display several 
sharp  
dips in correspondence with the peaks 
of the adatom DOS, cf. Fig.~\ref{fig_pdos}. Here, 
the transmission drops from 
the ideal value of 2 to 1
since one of the channels, $s$ or $a$, gets suppressed due to 
destructive interference between 
pathways going straight along the nanotube and those passing through the adatom orbital 
of the same symmetry.
If a DOS peak for one spin polarization occurs very close to the Fermi energy, 
then the mean-field conductance
of two spin channels differs significantly. 
That 
is the case, for example, for hollow site Co on the (8,8) nanotube and hollow site Fe on the (4,4) nanotube. Of course, these DFT results are expected to be
significantly altered by many-body effects in the low temperature regime (see discussion below).

The crucial 
quantities characterizing the scattering 
of conduction band states on the impurity 
are the scattering phase shifts. They are
obtained by diagonalizing the unitary $S$ matrix relating the amplitudes of outgoing and incoming scattering waves. 
In our case of a nanotube with two bands at the Fermi energy, the $S$ matrix (for each spin channel) will be a $(4\times4)$ matrix, 
two states provided by the left half of the nanotube and two by the right one. 
Figure~\ref{fig_sketch} shows schematically how the phase shifts are calculated. 
Let us consider for example the case of the hollow adsorption site. Here, by symmetry, 
the $s$ and $a$ channels do not mix so that
the $S$ matrix factorizes into two independent $(2\times2)$ blocks.  
We define as reference system a clean nanotube without the impurity, so that the 
unperturbed $S$ matrix is just the unit matrix. 
When the impurity is introduced at the hollow side, $S$ transforms into  
\begin{equation}\label{s_matrix}
S = 
\left( 
\begin{array}{cc}
t & r \\
r & t
\end{array} 
\right), 
\end{equation}
where $t$ and $r$ are transmission and reflection amplitudes, respectively. 
The matrix is symmetric due to
the mirror $xy$ symmetry plane. Diagonalizing $S$ 
we obtain
\begin{equation}\label{s_diagonal}
U^\dagger S U = 
\left( 
\begin{array}{cc}
e^{2i\delta_e} & 0 \\
0 & e^{2i\delta_o}
\end{array} 
\right),~~ 
U=\frac{1}{\sqrt{2}}
\left( 
\begin{array}{cc}
1 & 1 \\
1 & -1
\end{array} 
\right), 
\end{equation}
which reveals the two phase shifts, $\delta_e$ and $\delta_o$, corresponding to
even and odd eigenchannels as given by the columns of the unitary matrix $U$.
For each symmetry channel, $s$ or $a$, we thus obtain two phase shifts, even and odd.
From Eqs.~(\ref{s_matrix}) and (\ref{s_diagonal}) one can easily 
verify the following 
well-known relationship
between phase shifts and transmission and reflection probabilities:
\begin{equation}
|t|^2 = \cos^2(\delta_e-\delta_o),~|r|^2 = \sin^2(\delta_e-\delta_o) .
\label{tr_from_eo}
\end{equation}
On the other hand, the phase shifts can also be related to the extra DOS 
(of the same symmetry), $\Delta \rho$, induced by the impurity via the
Friedel sum rule:
\begin{equation}
\Delta \rho_i(E) = \frac{1}{\pi} \frac{d \delta_i(E)}{dE}, \hspace{10pt} i=e,o.
\label{sum_rule}
\end{equation}
The DFT phase shifts thus fully characterize the 
link between the electronic structure and the transport
properties of the system. We note that 
the two channels, $s$ and $a$, get mixed for the ontop adsorption site since
the symmetry plane $xz$
is missing, and one simply has two even and  
two odd phase shifts.

As an example, we present in Fig.~\ref{fig_tdelta} the transmission functions and phase
shifts for the case of hollow site Co on the (4,4) nanotube. Transmission functions 
for both spin channels and both symmetries, $s$ and $a$, are plotted on 
the upper panel while the even and odd phase shifts are shown on the middle and lower
panels, respectively. Since the phase shifts are defined modulo $\pi$, 
we choose to plot them on the interval $[-\pi/2,\pi/2]$.
One can see from the figure that all the dips in transmission are associated
with abrupt changes (by the value $\sim \pi$) in either even or odd phase 
shifts of the same symmetry,
in agreement with Eq.~(\ref{tr_from_eo}). 

These 
sharp features 
in the phase shifts 
are directly related to  
PDOS peaks of the same symmetry 
(see Fig.~\ref{fig_pdos}, upper panel), as 
implied by the Friedel sum rule, Eq.~(\ref{sum_rule}).  
For example, in spin down transmission of $s$ symmetry there are two dips 
at energies around 0.5 eV and 0.75 eV. The corresponding sharp features 
in the odd and even phase shifts derive, respectively, 
from the Co $d_{xz}$ and $s$ orbitals, 
hybridized with conduction electrons of the nanotube. 
The phase shifts calculated within the DFT approach 
are now ready to 
play the subsequent central role in 
generating
the parameters for the AIMs.

\section{Anderson models and their Hartree Fock phase shifts}\label{sec_anderson}
In this section we 
describe 
the method used to 
build 
effective AIMs starting from ab initio 
calculations of the nanotube with a transition metal impurity. 

The scattering calculations described in the previous section yield different phase shifts for 
spin up and spin down conduction electrons, which is of course an artifact of spin-rotational symmetry breaking.  
The root of the problem is that a broken-symmetry ab initio calculation misses quantum 
fluctuations between mean-field solutions with magnetization oriented in different directions, 
a process intrinsic to the Kondo effect that restores spin-rotational symmetry.  In the context of AIMs, 
a physically transparent way of starting from a mean-field solution with a pre-formed local moment and 
subsequently including quantum fluctuations is the Anderson-Yuval-Hamann path integral approach \cite{Hamann&PWA,Hamann}.   
In the same way, we could in principle restore spin symmetry by building quantum fluctuations on top 
of the ab initio calculation.  However, since our goal is to go from the ab initio data to the final result 
by way of a model Hamiltonian, we shall instead exploit the close analogy, mentioned in the introduction, 
between a spin-polarized DFT calculation and the mean-field solution of a generalized AIM.  Namely, 
we adjust the model parameters so that the scattering phase shifts of the AIM, at the mean-field level, 
exactly reproduce the ab initio phase shifts.  The model Hamiltonian thus obtained will provide a faithful 
low energy representation of the actual nanocontact if the quantum fluctuations in the model are in some 
sense similar to the local quantum fluctuations of the impurity.  All exact symmetries including spin-rotational 
symmetry are restored in the final step of our calculations, when the model Hamiltonian is solved with the NRG method.

Since the NRG method is numerically intensive, we will only be able to deal with a very limited number of channels and symmetries. 
Accordingly, our description of the electronic structure of the clean metallic 
tube will be necessarily crude, encompassing two 
conduction 
bands only. 
These bands are assumed to have a linear dispersion $E(k)=\pm v |k-k_F|$ around the Fermi energy (FE), 
giving rise to a constant density of states (DOS) $\rho_0=1/\pi v$ at the FE.  
As discussed in Sec.~\ref{sec_dft}, there are
four scattering channels (corresponding to the symmetries $es$, $ea$, $os$, $oa$), each with 
the same DOS at the FE. Each channel couples to the impurity orbitals with the same symmetry. 
It should be noted here that the neglect of all other nanotube subbands restricts our treatment
to SWNTs of smallest radius, where these subbands are sufficiently far from the Fermi level. This 
obstructs in particular any attempt to extrapolate towards the infinite radius limit, i.e.~graphene,
where all subbands coalesce at the FE.
%
The impurity orbitals that will be considered here are those in the valence shell 
of the TM atom, namely the five $3d$ orbitals and the $4s$ orbital, whose symmetry properties 
are listed in Table~\ref{tabsym}. 
When the impurity is in the \textit{hollow} 
site, the parities with respect to both reflection planes are good quantum numbers, 
and we can classify electronic states (both of the tube and the adatom) accordingly.  
When instead the impurity is in the \textit{ontop} position, only $e/o$ is a good quantum number, 
since symmetric and antisymmetric conduction states are mixed together. Here 
the problem is somewhat harder to treat; we will briefly illustrate this case later, while, in what follows, 
we shall always refer to the \textit{hollow} configuration, which is anyway the lowest in energy. 

Our general AIM includes therefore four scattering channels, $i=es,ea,os,oa$,  
and six impurity orbitals, $a=1,\dots,6$; hence, it is of the form
\bea
H &=& \sum_{ik\sigma}\,\Bigg(\epsilon_{k}\,c^\dagger_{ik\sigma}c^\dagga_{ik\sigma} + 
\sum_a\,V_{ik,a}\,\left(c^\dagger_{ik\sigma}d^\dagga_{a\sigma} + H.c.\right)\Bigg)\nonumber \\
&& + \sum_{ikk'\sigma}\,t_{i,kk'}\,c^\dagger_{ik\sigma}c^\dagga_{ik'\sigma} + H_{imp},\label{Ham-full}
\eea
where $c^\dagger_{ik\sigma}$ creates a spin $\sigma$ electron in channel $i$ with momentum $k$ along the tube, 
$d^\dagger_{a\sigma}$ a spin $\sigma$ electron in the orbital $a$ of the impurity. 
$V_{ik,a}$ is the hybridization matrix element between conduction and impurity orbitals, which is finite 
only if they share the same symmetry according to Table~\ref{tabsym}, while $t_{i,kk'}$ describes 
a local scalar potential felt by the conduction electrons because of the translational symmetry breaking 
caused by the impurity. $H_{imp}$ includes all terms that involve only the impurity orbitals, which we 
can write as 
\be
H_{imp} = \sum_{a\sigma}\, ( \epsilon_a\,n_a + U_a\,n_{a\up}\,n_{a\down})+\sum_{a<b}\,U_{ab}\,n_a\,n_b + H_{Hund},\label{H-imp}
\ee
where $n_{a\sigma}=d^\dagger_{a\sigma}d^\dagga_{a\sigma}$, $n_a=\sum_\sigma\,n_{a\sigma}$ and 
$H_{Hund}$ contains all interorbital interaction terms that in the isolated atom would give rise to 
Hund's rules. Since the degeneracy among the $d$ orbitals is fully removed in our 
scattering geometry, we will only take the first Hund's rule into account thus writing 
\be
H_{Hund} = \sum_{a< b}\, J_{ab}\,\mathbf{S}_a\cdot\mathbf{S}_b,\label{H-Hund}
\ee 
where $J_{ab}<0$, favoring a ferromagnetic correlation among the spin densities $\mathbf{S}_a$ of 
the different orbitals. 

The parameters in this Hamiltonian are so far unknown. As described earlier on, our goal is to establish a 
direct correspondence between a mean-field solution
of this model Hamiltonian, and the detailed DFT calculation of the previous chapter, 
that will allow, even if approximately, the extraction of ab initio based parameters.
The mean-field treatment of \eqn{Ham-full} is quite straightforward. One assumes that 
\ba
U_a\,n_{a\up}\,n_{a\down} &\longrightarrow& U_a\,\sum_\sigma\, \langle n_{a-\sigma}\rangle\,n_{a\sigma},\\
U_{ab}\,n_a\,n_b&\longrightarrow&U_{ab}\,\langle n_a\rangle\,n_b + U_{ab} \,\langle n_b\rangle\,n_a,\\
J_{ab}\,\mathbf{S}_a\cdot\mathbf{S}_b &\longrightarrow&  J_{ab}\,\langle \mathbf{S}_a\rangle \cdot\mathbf{S}_b
+ J_{ab}\,\langle \mathbf{S}_b\rangle \cdot\mathbf{S}_a,
\ea
where $\langle \dots \rangle$ is the average value, to be determined 
self-consistently, with respect to the Hartree-Fock Slater determinant. It follows that the Hartree-Fock 
Hamiltonian describes noninteracting orbitals, 
each one characterized by an effective spin-dependent energy 
\be\
\epsilon^{*}_{a\sigma} = \epsilon_a + U_a\,\langle n_{a-\sigma}\rangle \pm \frac{1}{4}\sum_{b}\,J_{ab}\,
\langle n_{b\up}-n_{b\down}\rangle+\sum_b\, U_{ab}\, n_b,
\label{epsilon_a-HF}
\ee
where the plus/minus sign refers to $\sigma=\up/\down$. Each channel $i$  
scattering off the impurity region acquires a spin-dependent phase shift $\delta_{i\sigma}$ 
caused by the potential term $t_{i,kk'}$ as well as the hybridization with the localized orbitals. 
We assume that $t_{i,kk'}$ alone would produce a scattering phase shift $\phi_i$. It follows that, 
if we concentrate on the region close to the chemical potential where the DOS is constant, 
the total phase shift satisfies the equation  
\be
\tan \delta_{i\sigma} = \tan \phi_i + \sum_a\, \fract{\Gamma_{ia}}{\epsilon^{*}_{a\sigma}},\label{delta_i}
\ee
where 
\[
\Gamma_{ia} = \pi\sum_k\,V_{ik,a}^2\,\delta\left(\epsilon_k-\epsilon_F\right)
\]
is the hybridization width at the Fermi energy. The ab initio knowledge 
of the phase shifts $\delta_{i\up}$ and $\delta_{i\down}$ allows us to 
fix only two parameters in Eq.~\eqn{delta_i}. 

When the channel $i$ is coupled to a single orbital, 
one could fix $\phi_i$ and $\Gamma_{ia}$ should $\epsilon^*_{a\sigma}$ be known.        
If the ab initio PDOS of the impurity orbital $a$ with spin $\sigma$ has a well pronounced peak at some 
energy, it is reasonable to identify the latter with $\epsilon^*_{a\sigma}$.  
This is generally the case, however, in some instances the PDOS of the impurity orbital has a long tail that extends up to 
the edge of the lowest subband, where not only the conduction electron DOS deviates strongly from the constant FE value $\rho_0$,
displaying a characteristic one-dimensional Van Hove singularity, but other 
subbands also contribute to the hybridization.
In such situations, the assumptions underlying Eq.~\eqn{delta_i} are no longer valid, and one should in 
principle take into account the energy dependence of the phase shifts and not just their value at the chemical potential. 
This is 
feasible but makes the calculations 
much more involved.  Instead, we adopted a simplified route consisting of keeping just 
the lowest subband, assuming 
a constant DOS $\rho_0$ and fixing $\epsilon^*_{a\sigma}$  
as the energy where the integrated PDOS is about one half. This assumption is justified only 
so long as 
the final results do not depend strongly 
on the precise choice of $\epsilon^*_{a\sigma}$, 
which we will verify {\sl a posteriori}. 

Having fixed $\epsilon^*_{a\sigma}$,
$\phi_i$ and $\Gamma_{ia}$, 
we now need 
to determine $\epsilon_a$, $U_a$ and $J_{ab}$ -- still too many parameters.
One can reduce them by assuming that $J_{ab}$ is constant within the 
$d$ shell ($J_{ab}=J$) and that $J_{sd}$ is the same for all $d$ orbitals. 
Another reasonable assumption, which can be verified directly in the ab initio calculation, 
is that the magnetization $M_s$ of the $s$ orbital is negligible, so that its spin splitting 
is controlled by the total $d$ magnetization $M_d$ through $J_{sd} M_d$, see Eq.~\eqn{epsilon_a-HF}.  
This fixes $J_{sd}$.  Then, $J$ can be determined through the spin-splitting $J M_d + J_{sd} M_s \simeq J M_d$ of the fully occupied/empty $d$ orbitals. 
The knowledge of 
$J$ and $J_{sd}$ allows us to determine $U_a$ of the partially 
filled $d$ orbitals. 
Finally, we fix $U_{ab}$ by
\be\label{u54j}
U_{ab}=U_{av}-\frac{5}{4}\,J,
\ee
where $U_{av}$ is the average of all $U_a$.  Equation~(\ref{u54j}) holds for an isolated atom \cite{liechtenstein_u54j}; 
we assume it remains approximately valid when the degeneracy of the $d$ orbitals is broken, since it involves an average.

We emphasize that $M_s$, $M_d$, and $\langle n_{a\sigma}\rangle$ as well as 
$\phi_i$ and $\Gamma_{ia}$ depend 
implicitly on the various parameters $J$, $J_{sd}$, $U_a$ and $U_{ab}$, so that fixing them actually requires 
solving the Hartree-Fock equations self-consistently. Once this program has been accomplished, 
all AIM parameters are determined in such a way that the mean field reproduces the ab initio 
phase shifts and the energetic position of the impurity levels.     

The above scheme works when each channel $i$ is coupled to a single orbital $a$. However,  
for Fe on the (4,4) nanotube the $es$ channel is coupled to two orbitals, 
$d_{z^2}$ and $s$. In this case further assumptions are required to determine the AIM 
parameters, which 
we 
shall 
discuss later. 

The AIM Hamiltonian \eqn{Ham-full} 
constructed 
in this way, already greatly 
simplified with respect to the full physical situation represented by the ab initio 
starting point,
still has too many degrees of freedom to be treated by accurate many-body techniques such as NRG. 
Since our final goal is to describe the low temperature and low bias properties, 
we can neglect orbitals that are either fully occupied or empty within DFT, provided 
the energy scale relevant for magnetic quantum fluctuations, i.e. the Kondo temperature, 
is much smaller than the energy required to excite electrons/holes from those orbitals. This condition 
has to be verified {\sl a posteriori}, but we anticipate that it actually holds.     
Discarding such inert orbitals, namely assuming that they are decoupled from the conduction electrons and just contribute to the scalar potential 
$t_{i,kk'}$ in Eq.~\eqn{Ham-full}, 
it turns out that the number of active orbitals is two for a Co impurity 
(one of them, $d_{xz}$, being 
half filled and 
magnetic and the other one, $d_{xy}$, almost filled). 
The number of relevant orbitals is instead  
three for Fe on an (8,8) tube  where $d_{xz}$ and $d_{xy}$ are magnetic, and 
$d_{z^2}$ almost filled,  and four in the case of Fe on the (4,4) tube, where besides the three orbitals 
of the (8,8) case also the $4s$ orbital is found 
to be partially occupied in DFT. 
In conclusion, for Co only the $os$ and $ea$ channels 
are effectively hybridized with the impurity orbitals $d_{xz}$ and $d_{xy}$, respectively. 
In the case of Fe, we must additionally include the hybridization between the $es$ channel and 
the $d_{z^2}$ orbital in the case of the (8,8) tube, and the $d_{z^2}$ and $s$ orbitals in the case of the (4,4) tube. 

\begin{table}\centering
\begin{tabular}{|c|c|c|}
\multicolumn{3}{c}{Hollow}\\
\hline & s&a\\
\hline e&$d_{z^2}$,$d_{x^2-y^2}$,$s$&$d_{xy}$\\
\hline o&$d_{xz}$&$d_{yz}$ \\
\hline
\end{tabular}
\quad
\begin{tabular}{|c|c|}
\multicolumn{2}{c}{Ontop}\\
\hline&\\
\hline e&$d_{z^2}$,$d_{x^2-y^2}$,$d_{xy}$,$s$\\
\hline o&$d_{xz}$,$d_{yz}$\\
\hline
\end{tabular}
\caption{Valence orbitals of Fe and Co and their symmetries ($e$/$o$: even-odd with respect to the $xy$ plane; $s$/$a$: symmetric-antisymmetric with respect to the $xz$ plane; $z$ is the axis of the tube). On the left \textit{hollow} configuration, on the right \textit{ontop} configuration.} \label{tabsym}
\end{table}

\section{Co inside and outside nanotubes: results}\label{sec_results}
In the previous section, we showed how to derive 
Anderson impurity models that should correctly capture the 
low temperature nanotube transport properties. We refer to appendix \ref{sec_nrg} for details 
about how to solve these models, and to appendix \ref{app_tables} for all DFT-GGA quantities
relevant for the different cases. 
All AIM parameters are listed in Table \ref{tabres}.  
In this section, we present the actual 
solution in the case of a Co impurity absorbed inside or outside a nanotube. 
This case will
also serve to explore and expose the possible magnitude of errors introduced by the inaccuracies of the
starting DFT electronic structure, generally attributed to incomplete cancellation of self-interactions.
It was found that these errors may be important in delicate cases where different orbitals compete,
calling for additional care at the outset of the calculations.     


\subsection{Co outside a (4,4) tube, 
hollow site 
}\label{co44}

According to DFT, in this geometry Co is in a configuration 
very close to $3d^94s^0$, and hence with spin $S=1/2$. In particular it has 
only one truly magnetic orbital, $d_{xz}$, coupled to the $os$ 
conduction 
channel, 
along with the
almost fully occupied, i.e.~only partially magnetized, 
$d_{xy}$ orbital coupled to the $ea$ channel.  
All other orbitals are assumed to be inactive, and
the effective AIM thus comprises two orbitals, 
each coupled to its own separate channel. The two impurity states however are 
coupled to one another by a ferromagnetic exchange $J$ and an interorbital Hubbard repulsion $U_{12}$. 
The 
two 
remaining channels $oa$ and $es$ are free (apart from  potential scattering)
since they do not couple to any magnetic orbital.  
Because it is somewhat unusual, the $S=1/2$ spin state of nanotube-adsorbed Co 
required some checking, to avert the possibility that it might arise as an artifact of,
for instance, GGA self-interaction errors. 
We found in fact that $S=1/2$ for Co on (4,4) is stable against removal of self interaction.
Using for example GGA+$U$ with $U$=2 eV, we obtained qualitatively the same result 
as for pure GGA: orbital $d_{xz}$ is magnetic, orbital $d_{xy}$ is almost 
fully occupied,
and orbital $s$ is empty. All relevant parameters are listed in Tables \ref{tabres} and \ref{co4}.

First of all, we performed an NRG run for each active channel 
ignoring their mutual coupling, 
that is setting $J=U_{12}=0$. In this way we found 
the phase shifts $\delta_{os}$ and $\delta_{ea}$ 
(indicated as $\delta_{NRG}$ in Table~\ref{tabres}) and, together with 
$\delta_{es}\equiv \phi_{es}$ and $\delta_{oa}\equiv{\phi_{oa}}$, 
the zero-bias conductance for each channel using 
Eq.~(\ref{gcos}) ($g_{NRG}$ in Table~\ref{tabres}). It turns out that this 
first-run phase shift is almost $\pi/2$ for the $os$ channel, 
as expected for a Kondo channel close to particle-hole symmetry, 
while the $ea$ channel suffers 
only 
a negligible phase shift, 
as the $d_{xy}$ orbital is 
almost fully occupied and potential scattering is negligible.
We also estimated a first-run Kondo temperature $T_K$ of the order of 3 K for the Kondo channel $os$.

We then performed 
a successive 
NRG run with both channels, now coupled by  $J$ and $U_{12}$.
The Kondo temperature 
decreased to 
$T_K\simeq 0.6$ K (this is indicated as $T_{K}$ in Table \ref{tabres}). 
The decrease is appreciable although not dramatic 
since 
$d_{xy}$ is almost fully occupied.
We now have a concrete example where we can check to what extent the errors implict
in the DFT starting point influence the calculation. 
The addition of a $U=2$ eV term within GGA+$U$ has the main result 
of increasing $|\epsilon_i|$ 
and $U_i$, while $\Gamma_i$ does not change appreciably. 
While this has no effect on the 
zero-temperature value of the zero-bias conductance, it leads to a 
strong decrease of $T_K$ well below 1 $K$, 
an inevitable outcome since $T_K$ depends on $U_i$ exponentially. 
The actual choice of the parameter $U$ in GGA+$U$ strongly influences the estimate 
of $T_K$, 
even though it has little effect on the electronic structure especially above a certain 
value. That is disappointing since there is no rigorous prescription 
for choosing the value of $U$. 
The apparent increase of the magnetic spin splitting and of $U_i$ upon increasing the parameter 
$U$ in GGA+$U$ is in fact 
more an artifact than a true physical effect. In fact, once $U$ has had its main role of 
pushing atomic occupancies closer to integer values, the physics becomes independent of $U$,
while the GGA+$U$ apparent spin splitting keeps on increasing artificially.
This is clearly a point that will require further work.
For this reason, we 
decided to determine 
$T_K$ through the parameters obtained by 
GGA without $U$ with the understanding that this will probably provide an upper estimate.  



\subsection{Co inside a (4,4) tube, hollow site}

The equilibrium configuration of Co inside the (4,4) SWNT 
mirrors the configuration outside
and has the same active orbitals $d_{xz}$ and $d_{xy}$.
However, in this case there is a switch of symmetry relative to the outside case. The  $d_{xy}$ 
orbital, essentially inactive outside,  is now much more hybridized with the nanotube, 
hence it loses charge and comes closer to being half-filled.
The charge is transferred partly to the $d_{xz}$, now $\sim 70\%$ occupied, 
and partly to the nanotube.
As a result, Co inside the nanotube is still a $S=1/2$ impurity as it was 
outside, but the magnetization is shared by both orbitals $d_{xz}$ and $d_{xy}$. 
The parameters that characterize the orbitals are listed in Tables \ref{tabres} and \ref{co4in}. 
%
%

Running NRG 
for these two orbitals coupled by a Coulomb repulsion $U_{12}$ 
as well as a ferromagnetic exchange $J$, we find a ground state configuration 
where $d_{xy}$ is close to half filled 
and $d_{xz}$ is close to fully occupied. This $S=1/2$ configuration 
for Co inside the nanotube 
still leads 
to a zero-bias conductance $G=G_0$, 
similar to 
Co outside, with the major 
difference that the Kondo temperature is now much larger---and the corresponding anomaly in the spectral 
function is much broader---since inside the orbital $d_{xy}$ is substantially more hybridized
than $d_{xz}$ was outside. The switching between $d_{xz}$ and 
$d_{xy}$ orbitals is mostly a geometrical effect and produces a much stronger hybridization of 
$d_{xy}$ with the nanotube.  The resulting increased delocalization of the 
$d_{xy}$ orbital %
implies that the value of $U_{12}$ 
obtained from fitting the Hartree-Fock mean field is must be somewhat lower than the estimate based on Eq.~(\ref{u54j}).  

Upon repeating the DFT calculation with GGA+$U$
($U=2$ eV), however, the orbital $d_{xz}$ became almost completely spin polarized, 
while orbital $d_{xy}$, being delocalized into the nanotube, is unaffected and remains
only modestly spin polarized. This is not
unexpected, since at the mean-field level the least hybridized orbital  generally becomes 
strongly magnetic.  
NRG shows  
that in this case the orbital $d_{xz}$ goes into the Kondo regime with a
low value of $T_K\sim 10^{-5}K$, while orbital $d_{xy}$ moves below the Fermi energy, 
is about 70\% filled and yields an appreciable decrease of the zero-bias conductance.
Thus, suppression of self interactions by inclusion of a Hubbard repulsion in the GGA calculation does not 
change the spin of the Co impurity, which remains always $S=1/2$, 
but may cause the orbitals to revert back to the case outside the tube, thereby lowering the Kondo temperature.
The persistence of a $S=1/2$ state 
contrasts with the case of Co/graphene \cite{wehling}, where GGA yields $S=1/2$, 
but GGA+$U$ favors the experimentally relevant $S=1$ configuration \cite{dubout}. The reason is that the $d_{xz}$ and 
$d_{xy}$ orbitals are degenerate on graphene due to the higher  $C_{6v}$ symmetry 
as opposed 
to the $C_{2v}$ symmetry of the nanotube 
hollow site. In the $3d^94s^0$ configuration given by GGA for Co on graphene, 
the minority-spin $d_{xz}$ and $d_{xy}$ orbitals lie exactly at the Fermi energy; this is an unstable situation 
when a Hubbard $U$ is added. It turns out that in this case the minority spin doublet moves above the Fermi 
energy, and charge neutrality is maintained by partially filling the $4s$ orbital, leading to 
a $3d^84s^{0.5}$ configuration with spin $S$=1. On the (4,4) nanotube, instead, the crystal field removes 
the degeneracy of the doublet in such a way that an integer occupation of both orbitals 
can be achieved already for $U=0$. 
In conclusion, it is likely that a transition $S=1\rightarrow S=1/2$ 
occurs in going from Co/graphene (or large nanotubes) to Co/small single wall nanotubes.
As noted earlier~\cite{baruselli_prl2012}, in small nanotubes it makes a qualitative difference within GGA whether the impurity 
is adsorbed inside or outside. 
For Co outside, the orbital $d_{xz}$ is in the Kondo regime with a small Kondo temperature. 
For Co  inside, the Kondo orbital is $d_{xy}$, whose hybridization is substantially larger because of the 
curvature, hence leading to a larger Kondo temperature inside as opposed to outside. 
However, if GGA+$U$ is to be trusted, the Kondo orbital would remain $d_{xz}$ in both cases, leading to similarly small 
Kondo temperatures inside and outside.  


\subsection{Co outside a (4,4) tube, ontop site}

In this geometry, the electronic configuration of Co in DFT 
is the same as it was in the hollow configuration, $3d^94s^0$, 
with active orbitals $d_{yz}$ and $d_{xy}$ (see Tables \ref{tabres} and \ref{co4top}). 
However, because of the lower 
symmetry, the $s$ and $a$ bands are mixed. 
In this case we need to use 
a more general expression for the conductance
\begin{eqnarray*}
 &G=\left[\cos^2(\delta_{es}-\delta_{os})+\cos^2(\delta_{ea}-\delta_{oa})\right]\cos^2(\theta_e-\theta_o)+\nonumber\\
&+\left[\cos^2(\delta_{es}-\delta_{oa})+\cos^2(\delta_{ea}-\delta_{os})\right]\sin^2(\theta_e-\theta_o),
\end{eqnarray*}
where $\theta_{e,o}$ are the mixing angles between $es$ - $ea$ and 
$os$ - $oa$ channels, which 
can be estimated from DFT calculations. 
However, in DFT these angles depend on the spin 
polarization 
because spin up 
and down are 
widely 
split and probe different energy regions. This would force 
us to use a more complicated model than the simple Anderson 
Hamiltonian (\ref{Ham-full}) 
with $k$-independent matrix elements $V_{ika}\equiv V_{ia}$. To simplify the analysis, 
we decided to drop the $d_{xy}$ orbital, whose effect is presumably small. 
This 
is equivalent to assuming that a particular linear combination of $s$ and $a$ 
bands is coupled to $d_{yz}$, 
while the orthogonal combination is free and gives unitary conductance. 
Within 
this approximation, we find that the results 
are similar to the hollow configuration as far as conductance ($\sim G_0$) but indicate a slightly larger Kondo temperature ($\sim 3$ K).


\subsection{Co outside an (8,8) tube, hollow site}
The configuration of Co on the (8,8) SWNT resembles that for the (4,4) SWNT and has the same active orbitals $d_{xz}$ (magnetic) and $d_{xy}$ (almost fully 
occupied). Therefore, the behavior of Co on the (8,8) SWNT is similar to 
that on the (4,4) SWNT (see Tables \ref{tabres} and \ref{co8}).  
Both $U$ and $\Gamma$ are slightly smaller than in the (4,4) tube, 
implying roughly the same conductance (close to $G_0$) and a somewhat smaller Kondo temperature 
(about 0.1 K when taking into account $J$ and $U_{12}$). It follows that 
the conductance does not depend appreciably 
on the size of the nanotube, provided it is small.


\subsection{Co inside an (8,8) tube, hollow site}		
The configuration of Co inside the (8,8) SWNT is similar 
to that of Co inside the (4,4) SWNT 
and has the same active orbitals $d_{xz}$ and $d_{xy}$.
However, in this case the latter orbital is less hybridized with the 
nanotube due to the reduced curvature. The various 
parameters that characterize the orbitals are listed in Tables \ref{tabres} and \ref{co8in}.

\section{Fe inside and outside nanotubes: results}\label{sec_fe}

After the above exhaustive study of the Co impurity, it is instructive to 
compare results with 
a Fe impurity,
which highlights some common aspects as well as differences.
 
\subsection{Fe outside a (4,4) tube, hollow site}
In the adsorbed Fe impurity, 
the orbitals $d_{xz}$ and $d_{xy}$ are both magnetic. 
In addition, GGA predicts that $d_{z^2}$ is also close to being magnetic 
and that $4s$ is partly occupied and polarized, leading to a fractional 
total magnetic moment $\mu=3.40\mu_B$. 
Therefore, unlike all previous examples, GGA results are here compatible with a mixture of $3d^84s^0$ 
($S=1$) and $3d^74s^1$ ($S=2)$. 

The AIM hence involves a total of four orbitals  
coupled to three channels, two of them ($4s$ and $3d_{z^2}$) 
with 
the same $es$ symmetry. 

The effective AIM for Fe is thus 
much more 
complicated since four orbitals are involved. 
In addition to three $d$ orbitals, the 4$s$ orbital is also partly occupied, its spin-up component being
exactly at the Fermi energy, so that the choice of genuine 
magnetic orbitals is not straightforward.
Some orbitals 
become magnetic, i.e.~partially filled, only in response to the magnetization 
of other orbitals. Since we know that spin symmetry 
must be recovered in the ground state, 
these orbitals 
should in the true ground state 
end up 
fully empty or fully occupied and
hence not Kondo active. 
Whereas in all the previous examples the distinction 
between genuine magnetic orbitals and 
orbitals that magnetize indirectly was 
clear, 
in the case of Fe~(4,4) there are 
uncertainties, especially regarding the orbitals $3d_{z^2}$ and $4s$.
If we just focus on these 
two, we need to solve an AIM with two 
nondegenerate orbitals hybridized to a single conduction channel and coupled  
to each other by a ferromagnetic exchange. Using the parameters extracted from GGA, 
we 
ran an NRG calculation for this model and found that in the ground state the $4s$ 
orbital is practically empty and the $3d_{z^2}$ fully occupied. Therefore we 
expect that, in contrast to 
the GGA starting point, the actual atomic configuration 
of Fe on the $(4,4)$ SWNT will be $3d^8\,4s^0$, which is the same we will find for the $(8,8)$ tube, 
with two magnetic orbitals, $d_{xz}$ and $d_{xy}$, and hence spin 
$S=1$. The NRG calculation for these two orbitals, each hybridized to a 
conduction channel and mutually coupled by Hund's rule ferromagnetic exchange, yields full 
Kondo screening with each channel acquiring a phase shift close to $\pi/2$. 
Small deviations from $\pi/2$ are caused by imperfect particle-hole symmetry. 
The final result is that the conductance at low bias and low temperature 
is pushed down to $G\simeq 0$.

However, we cannot rule out the occurrence 
in Fe 
of an $S=3/2$ state with the orbital $3d_{z^2}$ magnetic 
in addition to orbitals $3d_{xz}$ and $3d_{xy}$. That case is beyond our numerical capabilities, 
requiring three screening channels in the NRG calculation, but would most likely lead to a very 
low Kondo temperature and a zero-bias conductance of $G\sim G_0$. In this case, anisotropy 
would probably prevail and destroy the Kondo effect.

Results are summarized in Tables \ref{tabres} and \ref{fe4}.


\subsection{Fe outside an (8,8) tube, hollow site}

In this configuration Fe is close to $3d^84s^0$, 
thus carrying a magnetic moment $S=1$. 
As in the (4,4) tube, there are two magnetic orbitals, $d_{xy}$, 
coupled to the $ea$ channel and $d_{xz}$, 
coupled to the $os$ channel, and an almost fully occupied orbital, 
$d_{z^2}$, coupled to the $es$ channel.
The other orbitals can be safely assumed to be inactive, 
so the AIM comprises three orbitals, each coupled to 
a different conduction channel 
and coupled among themselves by a ferromagnetic exchange $J$ and a Coulomb repulsion $U_{12}$ (see Tables \ref{tabres} 
and \ref{fe8}).

Among the active orbitals, $d_{xz}$, $d_{z^2}$ and $d_{xy}$, the latter 
couples to the antisymmetric band and pushes the conductance down to 
zero in that channel.
The orbitals $d_{xz}$ and $d_{z^2}$ both couple to the symmetric band; 
the former causes a 
phase shift of about $\pi/2$ in the odd channel, the latter a phase shift 
close to zero in the even channel. 
It follows that the total conductance is nearly vanishing.
Since there are two magnetic orbitals, 
the system should exhibit two Kondo temperatures. 
However, it turns out that these are even lower than in the 
case of Co, and well below 1 K, 
due to the ferromagnetic Hund exchange $J$ between the two channels in Fe. 
Such low values of the Kondo temperature mean that other effects, for example 
spin anisotropy, which has been neglected in our work, will prevail and destroy the Kondo effect.
Results are summarized in Table~\ref{fe8}.

\subsection{Fe inside a (4,4) tube, hollow site}
The configuration of Fe inside the (4,4) SWNT is similar 
to the configuration outside  
and has the same active orbitals $d_{xz}$ and $d_{xy}$. However, $d_{z^2}$ is now predicted to be 
almost completely filled and the $4s$ orbital is empty and far from the Fermi energy, giving a total magnetic 
moment $\mu=1.84 \mu_B$.  As a consequence, the configuration is $3d^84s^0$ and carries a 
spin $S=1$, roughly the same as Fe on an (8,8) tube.
Like in the case with Co inside a (4,4) tube, the $d_{xy}$ orbital is strongly hybridized with the 
nanotube, leading to a high Kondo temperature, on the order of 30 K. However, in this 
case GGA+$U$ does not qualitatively affect the result, both orbitals $d_{xz}$ and $d_{xy}$ being always magnetic. The various 
parameters that characterize the orbitals are listed in Tables \ref{tabres} and \ref{fe4in}.

\subsection{Fe inside an (8,8) tube in the hollow position}
The configuration of Fe inside the (8,8) SWNT is similar 
to the one of Fe outside the (8,8) SWNT and has 
the same active orbitals $d_{xz}$ and $d_{xy}$. However, the hybridization of the $d_{xy}$ orbital, and hence its Kondo temperature, is now smaller due to the reduced curvature. The various 
parameters that characterize the orbitals are listed in Tables \ref{tabres} and \ref{fe8in}. 
 
Summing up, the Fe impurity is a multi-channel Kondo system, as opposed to a single channel for Co, but 
otherwise similar to Co. 
The highest predicted Kondo temperatures lie somewhat
below those expected for Co, owing to Hund's rule ferromagnetic exchange among the two channels.  

\begin{table*}[tbp]
$$
\begin{array}{|c|c|c|c|c|c|c|c|c|c|c|c|}
\hline
\mbox{System}&\mbox{Orb.}&\mbox{Sym.}&\Gamma&  \epsilon &U&J&U_{12}&T_{K}(K)&-\delta_{NRG}&g_{NRG}&g_{DFT}\\\hline



\hline
\mbox{Co/(4,4)}& d_{xz} &os   &0.087& -3.91 &2.42 &1.21&1.08&0.6&0.45	\pi&0.03&0.95\\
& d_{xy} &ea  &0.051& -4.34 &2.77 &''  &  ''  &  &0.11\pi&0.93&0.72\\

\hline
\mbox{Co/(4,4)}& d_{xz} &os   &0.087     & -6.02 &3.67 &1.27 &2.09 &0.001&0.50\pi&0.00&0.99\\
\mbox{U=2}   & d_{xy} &ea   &0.059& -6.46 &3.69 &''   &''      &  &0.05\pi &0.98&0.90\\
\hline

\mbox{Co/(4,4)}& d_{yz}&o &0.076&-3.41 &2.43&1.25&0.75&3&0.50\pi&0.01&0.99\\
\mbox{ontop}& d_{xy}&e &0.059&-3.54 &2.20&''&''&&&1&0.78\\
\hline

\mbox{Co/(4,4)}& d_{xz} &os&0.079     & -2.62 &2.37 &1.07&0.15& &0.22\pi&0.59&0.49\\
\mbox{inside}& d_{xy} &ea&    0.380  & -2.19 &1.99 &''  &''&    600  &0.41\pi&0.14&0.46\\
\hline

\mbox{Co/(4,4)}& d_{xz} &os&   0.086  &-6.52  &4.73 &1.44&1.62&10^{-5} &0.50\pi&0.01&0.88\\
\mbox{inside U=2}& d_{xy} &ea&  0.365   & -4.10 & 2.11&''  &''&     &0.28\pi&0.38&0.51\\
\hline


\mbox{Co/(8,8)}& d_{xz} &os&0.058     & -3.98 &2.49 &1.15&1.02& 0.1&0.45\pi&0.04&0.95\\
& d_{xy} &ea&0.038                 & -3.98 &2.43 &''  &'' &     &0.09\pi&1.00&0.53\\
\hline

\mbox{Co/(8,8)}& d_{xz} &os&0.080     & -4.30&2.55 &1.16&1.13&&0.10\pi&0.90&0.66\\
\mbox{inside}& d_{xy} &ea&0.126                 &-4.22 & 2.60&''  &'' &  25  &0.43\pi&0.05&0.63\\
\hline


\mbox{Fe/(4,4)}&d_{xz} &os&0.092&-3.12&2.34&1.16&1.08&0.002&0.50\pi&&\\
& d_{xy} &ea&0.082&-3.94&2.72&''&''&0.3&0.41\pi&0.09&0.93\\
 &d_{z^2} &es&0.060&-2.63&2.23&''&&&0.03\pi&0.03&0.50\\
 &s &es&0.039&0.79&0.12&''&&&&&\\ 

\hline

\mbox{Fe/(4,4)} &d_{xz} &os&0.126& -2.49&2.41&1.14&0.71&0.05&0.50\pi&&\\
\mbox{inside} & d_{xy} &ea&0.396& -2.01&1.86&''&''&30&0.51\pi&0.00&0.76\\
 &d_{z^2} &es&0.363&-2.11&1.33&''&&&0.12\pi&0.16&0.56\\

\hline

\mbox{Fe/(8,8)} &d_{xz} &os&0.062 & -2.90 &2.42 &1.15  &1.00 &\sim 10^{-7} &0.50\pi&&\\
 &d_{xy} &ea&0.044 & -2.95 &2.45 &''  &'' & \sim 10^{-8}&0.50\pi&0.00&0.98\\
& d_{z^2} &es&0.057 &-2.37 &1.88 &'' & &                  &0.03\pi&0.01&0.69\\

\hline

\mbox{Fe/(8,8)} &d_{xz} &os&0.081 & -3.14 &2.56 &1.14  &1.18 &\sim 10^{-5} &0.49\pi&&\\
 \mbox{inside}&d_{xy} &ea&0.134 & -3.19 &2.65 &''  &'' &0.01&0.48\pi&0.01&0.90\\
& d_{z^2} &es&0.112 &-1.81 &1.02 &'' & &                  &0.04\pi&0.04&0.88\\

\hline

\mbox{(4,4) Long.} &\sigma &e&0.065 & -1.56 &2.60 & 0& 0&\sim 10^{-3} &0.51\pi&&1.00\\
 \mbox{Vacancy}&\pi &e&0.523 &  & &''  &'' &&&&0.65\\

\hline

\mbox{(4,4) Transv.} &\sigma &/&0.124 & -1.43 &2.53 &0  &0 &\sim 1 &0.51\pi&&0.97\\
 \mbox{Vacancy}&\pi &/&0.422 &  & & '' & '' &&&&0.54\\

\hline
\end{array}
$$
\caption{Recapitulative table of the parameters used in the Anderson Hamiltonian for different configurations of Fe and Co on both (4,4) and (8,8) tubes; for Co (4,4) in the hollow position (both outside and inside) a calculation with GGA+$U$ ($U=2$ eV) is also reported. For each case, we report the orbitals that are involved in transport and their symmetries, the Anderson parameters (the broadening $\Gamma$, the level energy $\epsilon$, the Hubbard repulsion $U$, the Hund exchange $J$ and the interorbital Hubbard repulsion $U_{12}$), the results of the solution of the Anderson model (the Kondo temperature $T_K$, the NRG phase shift $\delta_{NRG}$ and the conductance $g_{NRG}$ in units of $G_0$ for each band), and finally the DFT conductance per band at the Fermi energy $g_{DFT}=g_{DFT}^\uparrow +g_{DFT}^\downarrow$ for the purpose of comparing with the many-body result.
}\label{tabres}
\end{table*}

\section{Predicted Kondo zero-bias conductance anomalies}\label{sec_zba}
In the previous section, we discussed the Kondo temperature and the zero-bias conductance of Co and Fe impurities. 
Now we 
extend the discussion to finite bias effects, by means of the Keldysh method for nonequilibrium Green's functions \cite{meir}. 
The conductance for a single band can be expressed in the form of a Fano \cite{Fano} resonance
\begin{equation}\label{cond1band}
 g_{s,a}(v)\equiv\frac{G_{s,a}(v)}{G_0}=\frac{(q+v)^2}{(q^2+1)(v^2+1)}, \hspace{10pt}v\equiv \frac{-eV_B-\epsilon_K}{\Gamma_K},
\end{equation}
where $v$ is the dimensionless bias potential ($V_B$, $\Gamma_K$ and $\epsilon_K$ are respectively 
the bias potential, the width of the resonance, proportional to the Kondo temperature, and the energy of the 
Kondo peak, in eV); $q$ is the Fano parameter, which describes the shape of the ZBA: $q=0$ means 
an anti-Lorentzian shape, $q=\pm\infty$ a Lorentzian one, and $q=\pm 1$ gives rise to the most asymmetric lineshapes. 


The above formula Eq.~(\ref{cond1band}) holds for a single band.
If we assume no coupling between the two bands (that is, $J=U_{12}=0$), we can get the total conductance by
simply adding the results from each band:
\be
g_{tot}(v)=g_s(v)+g_a(v).\label{totcond}
\ee
This no longer strictly true if the bands are coupled to each other. However, since the treatment becomes quite 
involved in that case, we will simply assume that Eq.~(\ref{totcond}) still holds approximately. Results are shown in  Table \ref{fanotable}.







\begin{table}[ptb]
$$
\begin{array}{|c|c|c|c|c|c|}\hline
\mbox{Impurity}&\mbox{Nanotube}&\mbox{Position}&\mbox{Orbital}&\Gamma_K(eV)&q\\\hline
\mbox{Co}	&(4,4)	&\mbox{Out}&d_{xz}&	1.5\times 10^{-4}&	-0.03\\
\mbox{Co}	&(8,8)	&\mbox{Out}&d_{xz}&	2.5 \times 10^{-5}&	-0.04\\
\mbox{Co}	&(8,8)	&\mbox{In}
			&d_{xy}&6.3\times 10^{-3}	&-0.10	\\
\mbox{Co}	&(4,4)	&\mbox{In}
&d_{xy}&0.15	&	-0.11\\\hline
\mbox{Fe}	&(4,4)	&\mbox{Out}&d_{xz}&	5 \times 10^{-7}	&0.01\\
		&	&		&d_{xy}&8\times 10^{-5}	&	-0.02\\
\mbox{Fe}	&(8,8)	&\mbox{Out}&d_{xz}&	3\times 10^{-11}	&0.06\\
		&	&		&d_{xy}&3\times 10^{-12}	&	-0.02\\
\mbox{Fe}	&(8,8)	&\mbox{In}&d_{xz}&3\times 10^{-9}		&0.09\\
		&	&		&d_{xy}&4\times 10^{-6}	&	-0.01\\
\mbox{Fe}	&(4,4)	&\mbox{In}&d_{xz}&1.3\times 10^{-5}	&0.35\\
		&	&		&d_{xy}&8\times 10^{-3}	&-0.06	\\\hline
\end{array}
$$
\caption{Relevant parameters for the conductance at finite bias; for each case we show the magnetic orbital, the width of the Kondo resonance $\Gamma_K$ and the Fano parameters $q$ (GGA results are shown).
}\label{fanotable}
\end{table}

\section{Kondo effect of vacancies in carbon nanotubes}\label{sec_vac}
Single-atom vacancies in a nanotube represent a 
simpler and more intrinsic magnetic impurity  
than the adsorbed transition metal atom described in the previous sections. 
We apply the same method described for Co and Fe impurities to a single-atom vacancy in a (4,4) nanotube. 
However, due to the lower symmetry of this situation, calculations can only be pursued to a 
more modest 
degree of accuracy.

When removing a carbon atom from a graphene sheet, carbon nanotube or nanoribbon, a magnetic 
moment arises \cite{lehtinen_prl,yazyev_helm,  palacios_vacancy,yazyev} due to the breaking of 
three $\sigma$ ($2sp^2$ hybrid) bonds 
and one $\pi$ ($2p_z$) bond. The magnetic moment has 
been found by DFT to be close to $1\mu_B$ \cite{lehtinen_prl,yazyev_helm}, hinting at $S=1/2$, even 
though in some situations magnetism seems to disappear\cite{lehtinen_njp}. The magnetic moment, 
being embedded in a metal (armchair nanotube) or in a semi-metal (graphene), should give rise to 
a Kondo effect. The case of graphene has been investigated both theoretically \cite{sengupta,
cornaglia_kondo_graphene, vojta_graphene} and experimentally \cite{Chen2011}, but the vacancy 
Kondo effect has not been addressed so far in nanotubes. 

Out of the three dangling $\sigma$ bonds created by the vacancy, 
two are mutually saturated 
when the two respective C atoms come together forming a weak Jahn-Teller type bond. 
In general, three different pairs can form, corresponding to three different static distortions
of the carbons lying next to the vacancy. In graphene, these three configurations are equivalent, since 
they can be transformed into one another by rotating the whole system by $\pm2\pi/3$ given the 
local $C_{3v}$ symmetry. 
In an armchair nanotube, where only a symmetry $C_s$ is preserved, two out of three configurations are still 
equivalent, and we will call them ``transverse'' (T). The third configuration is inequivalent 
and we will call it ``longitudinal'' (L) (see Fig.~\ref{fig_struc}). In our (4,4) SWNT, a DFT calculation shows 
that the transverse configuration is energetically favorable over the longitudinal one by about 0.8 eV.


Nonetheless, we will 
consider both T and L cases to illustrate the differences that arise.
In all cases the JT relaxation saturates two $\sigma$ dangling bonds. Two remaining broken bonds 
are left unsaturated -- one $\sigma$ on the C atom (called $C_1$) which is left unpaired, and one $\pi$. 
In both T and L configurations, 
the 
DFT calculated 
magnetic moment is close to $1\mu_B$ and mainly carried by the $\sigma$ orbital localized on 
the lone atom $C_1$. In addition, a $\pi$ symmetry state appears just below the Fermi energy (see Fig.~\ref{fig_pdos}). 
The corresponding wavefunction is delocalized around the defect, indicating that 
unlike the sigma broken bond, which is localized, 
the broken $\pi$ bond undergoes strong
delocalization within the $\pi$ nanotube conduction band. 
The hybridization of the broken $\pi$ bond 
with the nanotube bands 
appears strong enough to inhibit the spontaneous formation of a full magnetic moment. The corresponding
electronic states exhibit 
a small spin splitting and some magnetization but only as a result of intra-atomic exchange with the strongly magnetized $\sigma$ broken bond.
We find that in the transverse configuration, the $\pi$ vacancy state 
is antiferromagnetically coupled to the $\sigma$ orbital spin, leading to a total magnetic moment smaller than $1\mu_B$. In the 
longitudinal configuration, instead, the coupling is very weakly ferromagnetic, yielding a total magnetic 
moment which is slightly larger than $1\mu_B$.
This difference can be traced to the fact that the correlations are ferromagnetic 
within a sublattice 
and antiferromagnetic between the two sublattices. The exchange coupling is ferromagnetic for two orbitals 
localized on the same atom due to Hund's rule, but 
antiferromagnetic for two orbitals on neighboring atoms, as in the Hubbard model.
The $\pi$ orbital is delocalized around the defect, and its exchange 
coupling with the $\sigma$ state will be ferromagnetic or antiferromagnetic according to its weight on the 
various C atoms, a property that is evidently controlled by geometry. 

In Fig.~\ref{fig_t}, we show the spin-polarized transmission for the two types of vacancies. One conduction channel is always decoupled from the $\sigma$ and $\pi$ impurity orbitals, giving a contribution $\sim G_0$ to the total conductance. Both up- and down-spin $\sigma$ orbitals being far from the Fermi energy, the DFT conductance at zero bias is dictated by the $\pi$ orbital. 
In the longitudinal configuration both up- and down-spin $\pi$ orbitals are about 0.7 eV below the Fermi energy, contributing $\sim 0.65 G_0$ to the total zero-bias conductance $G= 1.65 G_0$. In the transverse configuration, the up-spin orbital is about 0.2 eV below Fermi, leading to $G_{\up}=0.2 G_0/2$, while the down-spin orbital is 1.1 eV below Fermi, leading to $G_{\down}=0.9 G_0/2$, which together with the decoupled channel gives a total conductance of $G=1.5 G_0$. 





For simplicity, we keep just the $\sigma$ orbital in building the AIM, which leads to a one-channel Hamiltonian whose parameters are shown in Table \ref{tabres}.
 
The small $T_K$ that we find in both cases contrasts with the case of graphene, where the experimentally-determined $T_K$ is between 30 and 90 K \cite{Chen2011}. 
We tentatively attribute this 
difference 
to the large curvature of the nanotube, which should substantially modify the hybridization of the $\sigma$ orbital with conduction channels.


\section{Discussion and conclusions}\label{sec_conclusions}

We have shown in a detailed case study how one can combine {\sl ab initio} 
electronic structure 
calculations with numerical renormalization group to get 
quantitative 
estimates
of the Kondo temperature and zero bias anomaly in the transport across 
atomically, structurally and electronically controlled nanowires. 
The specific examples we have chosen to apply this strategy 
to are Co and Fe magnetic impurities adsorbed on single-wall 
carbon nanotubes and a carbon vacancy in a pristine nanotube. While there are no experimental data
for these systems, their extreme simplicity and reproducibility recommend
them as ideal test cases for future study. Even in the absence of experimental data, the 
effect of various approximations and DFT errors can be tested here rather
instructively.  

Our main results can be summarized 
very shortly.  A Co atom (or a C vacancy) behaves 
effectively as a $S=1/2$ impurity and reduces the 
zero-bias conductance from the ideal value of $G=2G_0=4e^2/h$ down to $G=G_0$. 
On the contrary, Fe is found to be $S=1$ and should be able to completely suppress 
the conductance, $G=0$. 
This reduction in the conductance takes place over a range of temperature/bias determined 
by the typical scale $T_K$ of Kondo screening. We generally estimate $T_K$ to be small, a fraction of a Kelvin, and
hence hard to detect. The only exception is Co or Fe inside the narrowest (4,4) tube, where 
the curvature leads to a substantial enhancement of the hybridization between the magnetic orbital 
$d_{xy}$ and the nanotube, pushing the Kondo temperature fairly high. Even this result
has some level of uncertainty, since we do not yet know if the self-interaction error in DFT is excessive or not. 
In all other cases, the tiny values of $T_K$ make our results
difficult to verify experimentally, which is a bit disappointing. There is 
however a positive implication, namely that, according to these results,   
the 
increase in the resistivity observed in nanotubes below 
100 $K$ \cite{hone,grigorian,fisher} as well as the associated peak in the 
thermopower \cite{hone, grigorian} could indeed be 
caused by magnetic impurities trapped inside the tube.  
A remarkable magnetic impurity is the single carbon atom vacancy. Although its Jahn Teller
distorted structure and $S$=1/2 should in principle resemble that of a vacancy in graphene, the predicted
Kondo temperature is substantially smaller in the nanotube, most likely owing to the lower hybridization
caused by curvature.
  
From a general methodological perspective, our study highlights  
several interesting elements and difficulties 
in the ab initio modeling of magnetic impurities in nanoscale conductors.   
The method we have presented combines the advantages of DFT and NRG in a simple and easily manageable way; the former allows us to identify the magnetic orbitals and their electronic properties, while the latter correctly incorporates the quantum fluctuations that restore spin symmetry. 
In this respect it could be quite 
effective in many other cases, e.g.~magnetic molecules. 
The most important difficulty is that the Anderson model parameters obtained from spin-polarized GGA,
where the Kohn-Sham energy levels are generally affected by unknown self-interaction errors,
are not always reliable and often need to be corrected, for example by means of  GGA+$U$,
as repeatedly stressed above.
The other important aspect that is worth emphasizing is the distinction 
between what we might call ``driving'' and ``driven'' magnetic orbitals. Since spin-polarized GGA  
breaks spin symmetry, there are orbitals that become partially 
polarized only in response to the full magnetization of 
other orbitals to which they are coupled by Coulomb (Hund's rule) exchange. The net effect is that within GGA 
one often finds a fractional magnetic moment, which cannot be 
straightforwardly associated with a definite-spin Kondo impurity.
On the other hand, since the actual ground state must be 
spin-rotationally invariant, it is likely that orbitals that appear weakly polarized 
in GGA are in reality either empty or fully occupied, and hence nonmagnetic.  
Only a better calculation that takes quantum fluctuations properly into account can 
clarify this issue, which is exactly what NRG is suited for. 
In conclusion, the combined DFT+NRG approach that 
we have exemplified is able to provide a satisfactory, if approximate, description of the 
actual low temperature and low bias behavior of a magnetic nanocontact. 
Nevertheless, we should stress that several quantitative aspects, 
like the precise width and lineshape of the Kondo anomaly, remain uncertain 
mainly because the actual magnitude of the Hubbard $U$ in the Anderson model depends on the 
way GGA is implemented and small changes of $U$ may lead to appreciable 
changes of $T_K$.

\begin{acknowledgments}
This work was largely supported by PRIN/COFIN 2010LLKJBX. It also benefitted by 
the environment created by EU-Japan Project LEMSUPER, Sinergia Contract CRSII2$_1$36287/1,  
and ERC Advanced Grant 320796 MODPHYSFRICT.
\end{acknowledgments}

\bibliographystyle{apsrev4-1}

%

\appendix
\section{Numerical renormalization group calculations}\label{sec_nrg}

In this appendix section we show 
how the impurity model of Eq. \ref{Ham-full}
is 
solved by means of 
numerical renormalization group calculations.

NRG is a numerical technique 
originally developed by K. G. Wilson \cite{wilson} 
to solve Anderson and Kondo impurity models. It is by now a well established impurity solver, whose 
technical details can be found in many review papers, see e.g. Ref.~\onlinecite{bulla08}.
The NRG method imposes a logarithmic discretization of the conduction band
controlled by a discretization parameter $\Lambda$.
After a sequence of transformations, the discretized
model is mapped onto a semi-infinite chain whose first site is
the impurity spin.  
The Hamiltonian of the chain is diagonalized
iteratively starting from the impurity site and
successively adding degrees of freedom to the chain.
The key parameters that control the accuracy of the calculations are 
(1) the discretization parameter $\Lambda$, which 
should be as close as possible to one; (2) the number of states that are kept at each iteration.  
In our computations we used $\Lambda=1.8$ and kept about 1500 states per iteration 
for runs with a single conduction channel, and $\Lambda=2.5$ and about 3500 states for runs 
with two conduction channels. 

The main purpose of the NRG calculations is to determine the low temperature 
conductance across the impurity and the energy scale $k_B T_K$, where $T_K$ is the Kondo temperature, that controls the 
asymptotic low temperature regime. 

The zero-temperature and zero-bias conductance $g\equiv G/G_0$ is fully determined by the phase shifts according to the formula
\begin{equation}\label{gcos}
g=\cos^2(\delta_{es}-\delta_{os})+\cos^2(\delta_{ea}-\delta_{oa})\equiv g_s+g_a,
\end{equation}
where 
we assume $\delta_{ij}=\phi_{ij}+\delta_{ij}^{NRG}$ ($i=e/o$, $j=s/a$). Here
the $\phi_{ij}$ are obtained by DFT and $\delta_{ij}^{NRG}$ are extracted by 
NRG neglecting the scalar potential $t_{i,kk'}$. 
This assumption is justified since $\phi_{ij}$ are quite small.  

In the clean tube, $G=2G_0$ because there are two conduction bands crossing the 
chemical potential.
In the presence of the impurity, the phase shift acquired by  
each channel $l$ can be estimated by NRG through 
\begin{equation}
 \delta^{NRG}_l=\pi\,\frac{E_{l1}}{E_{l2}-E_{l1}}, \hspace{10pt} l=es, ea, os, oa,
\end{equation}
where $E_{l1}$ and $E_{l2}$ are the first two eigenvalues in the subspace with quantum numbers that 
correspond to adding one more electron to channel $l$. As mentioned, 
no NRG calculation is required the channel is decoupled from the impurity orbitals, the phase shift being merely due to potential scattering, 
$\delta_l=\phi_l\simeq 0$.

An alternative but equivalent way to estimate $\delta^{NRG}_l$
is to use the real part of the self energy at 
zero frequency, $\Re\Sigma(0)$, extracted from the spectral function, since
\begin{equation}
 \delta^{NRG}_l=\arctan\frac{\Gamma_l^*}{\epsilon_l^*+\Re \Sigma_l(0)}.
\end{equation}
These two ways of estimating the phase shifts lead to similar results. Specifically, we always find   
$\delta_l \simeq \pi/2$ for Kondo channels (channels 
coupled to a magnetic impurity
state), 
and $\delta_l \simeq 0$ for channels coupled only to almost filled or empty orbitals.

We extracted the Kondo temperatures  
of the magnetic channels from the Matsubara self-energy
\begin{equation}
 \Sigma_l(i\epsilon)=G_{0l}^{-1}(i\epsilon)-G^{-1}_l(i\epsilon),
\end{equation}
where
\begin{equation}
 G_{0l}(i\epsilon)=\frac{1}{i\epsilon-\epsilon_l+i\Gamma_l}
\end{equation}
is the noninteracting Green's function (describing the impurity with 
$U_l$ and $J$ set to zero)  and
\begin{equation}
G_l(i\epsilon)=\frac{1}{i\epsilon-\epsilon_l-\Sigma_l(i\epsilon)+i\Gamma_l}=
\frac{1}{Z_{part}}\sum_{n}\frac{|\langle GS|d_l|n\rangle|^2}{i\epsilon-\epsilon_n}
\end{equation}
is the full Green's function ($Z_{part}$ is the partition function, $GS$ the ground state, and $d_l$ annihilates
an electron in the impurity orbital with symmetry $l$). 
Matrix elements $\langle n|d_l|m\rangle$ were collected during the NRG runs according to the 
patching technique \cite{bulla01}. This approach introduces small deviations with respect to more refined techniques such as those described in \onlinecite{peters_dmnrg,weich_dmnrg}, but, due to our large incertainties in the determination of parameters, it does not affect substantially our final results.

The Kondo temperature is given by
\begin{equation}
 T_K^l=\frac{\pi wZ_l\Gamma_l}{4k_B},
\end{equation}
where $w=0.4128$ is the Wilson coefficient and $Z_l$ is the quasiparticle residue
\begin{equation}
 Z_l^{-1}=1-\frac{\partial\Sigma_l(i\epsilon)}{\partial(i\epsilon)}.
\end{equation}
The
Kondo temperature is related to the width of the Kondo peak $\Gamma_K^l=Z_l \Gamma_l$ in the PDOS through
\begin{equation}
k_BT_K^l=\frac{w\pi}{4}\Gamma_K^l=0.342\Gamma_K^l.
\end{equation}




\section{Further numerical data}\label{app_tables}
In Tabs. \ref{co4} to \ref{fe8in} we present some further DFT data, and related extracted quantities, for different cases.

\begin{table*}[ptb]
 $$
\begin{array}{|c|c|c|c|c|c|c|c|c|c|c|}\hline
orbital&\epsilon_{\uparrow}&\epsilon_{\downarrow}&\delta_{\uparrow}&\delta_{\downarrow}&\Gamma_{PDOS}&\Gamma&\phi&\delta_{NRG}&g_{NRG}&g_{DFT}\\\hline
d_{x^2-y^2}(es)&-1.16&-0.36&0.118&0.023 &&0.050&0.160&&&\\
d_{xz}(os)     &-1.56&0.47 &0.133&0.358 &0.080	&0.087&0.187&-1.41&0.03(s)&0.95\\
d_{xy}(ea)     &-1.18&-0.04&0.156&-0.827&0.062&0.051&0.198&-0.35&&\\
d_{yz}(oa)     &-2.16&-1.36&0.016&-0.038&&0.197&0.107&&0.93(a)&0.72\\\hline
\end{array}
$$

\caption{Case of the Co/(4,4) system.  Relevant DFT quantities (energy of $d$ orbitals $\epsilon_{\sigma}$, in eV, and phase shifts $\delta_{sigma}$ for each spin direction), parameters extracted from the Hartree-Fock analysis (broadening of the level $\Gamma$, in eV, which is compared to the broadening directly extracted from the PDOS, $\Gamma_{PDOS}$, and the potential-scattering phase shift $\phi$, in radians), and results from the NRG runs (phase shift $\delta_{NRG}$, in radians, and conductance $g_{NRG}$, in units of $G_0$, compared to the DFT prediction $g_{DFT}$).}\label{co4}

\end{table*}

\begin{table*}[ptb]
 $$
\begin{array}{|c|c|c|c|c|c|c|c|c|c|c|}\hline
orbital&\epsilon_{\uparrow}&\epsilon_{\downarrow}&\delta_{\uparrow}&\delta_{\downarrow}&\Gamma_{PDOS}&\Gamma&\phi&\delta_{NRG}&g_{NRG}&g_{DFT}\\\hline
d_{x^2-y^2}(es)&-1.31&-0.86&-0.109&-0.321&    &0.392  & -0.210  &&&\\
d_{xz}(os)     &-1.39&0.03&0.053&1.274	&0.079&0.095	& 0.121  & -0.31  & 0.99(s)&0.49 \\
d_{xy}(ea)     &-0.71&-0.12&-0.309&-1.245&0.407&0.380&0.210&-1.29&&\\
d_{yz}(oa)     &-1.69&-1.19&0.081&0.073&       &0.031 &0.099 && 0.14(a)&0.46\\\hline
\end{array}
$$

\caption{Same as Table \ref{co4} for Co inside the (4,4) tube.}\label{co4in}
\end{table*}

\begin{table*}[ptb]
 $$
\begin{array}{|c|c|c|c|c|c|c|c|c|c|c|}\hline
orbital&\epsilon_{\uparrow}&\epsilon_{\downarrow}&\delta_{\uparrow}&\delta_{\downarrow}&\Gamma_{PDOS}&\Gamma&\phi&\delta_{NRG}&g_{NRG}&g_{DFT}\\\hline
d_{x^2-y^2}(e)&-1.28&-0.60&0.161&0.060&&0.116&0.248  &&&\\
d_{xz}(o)&-1.42&-0.84&0.028&-0.063&&0.187&0.159 & && \\
d_{xy}(e)&-1.15&-0.07&0.021&-0.655&&0.059&0.072&&1&0.79\\
d_{yz}(o)&-1.96&0.41&0.077&0.294&0.083&0.076&0.116&-1.48&0.02&0.99\\\hline
\end{array}
$$

\caption{Same as Table \ref{co4} for Co in the ontop position on the (4,4) tube. Due to the lower symmetry of this case, it is not possible to assign unequivocally an orbital to each channel; we just show the orbital with the highest weight coupled to each channel.}\label{co4top}
\end{table*}

\begin{table*}[ptb]
 $$
\begin{array}{|c|c|c|c|c|c|c|c|c|c|c|}\hline
orbital&\epsilon_{\uparrow}&\epsilon_{\downarrow}&\delta_{\uparrow}&\delta_{\downarrow}&\Gamma_{PDOS}&\Gamma&\phi&\delta_{NRG}&g_{NRG}&g_{DFT}\\\hline
d_{x^2-y^2}(es)&-1.52&-0.64&0.091&0.046&&0.051&0.125&&&\\
d_{xz}(os)&-1.68&0.26&0.129&0.365&0.057&0.058	&0.162&-1.41&0.04(s)&0.95\\
d_{xy}(ea)&-1.36&0.01&0.121&1.368&0.039&0.038&0.149&-0.28&&\\
d_{yz}(oa)&-2.00&-0.94&0.076&0.036&&0.071&0.111&&1.00(a)&0.53\\\hline
\end{array}
$$

\caption{Same as Table \ref{co4} for the Co/(8,8) system.}\label{co8}
\end{table*}

\begin{table*}[pt]
 $$
\begin{array}{|c|c|c|c|c|c|c|c|c|c|c|}\hline
orbital&\epsilon_{\uparrow}&\epsilon_{\downarrow}&\delta_{\uparrow}&\delta_{\downarrow}&\Gamma_{PDOS}&\Gamma&\phi&\delta_{NRG}&g_{NRG}&g_{DFT}\\\hline
d_{x^2-y^2}(es)&-1.68&-1.05&0.026&-0.037&    &0.178  & 0.131  &&&\\
d_{xz}(os)     &-1.63&0.07&0.068&0.921	&0.070&0.080	& 0.117  & -0.31  & 0.90(s)&0.66 \\
d_{xy}(ea)     &-1.37&0.08&0.049&1.063&0.099&0.126&0.140&-1.35&&\\
d_{yz}(oa)     &-1.69&-1.19&0.081&0.030&       &0.010 &0.137 && 0.05(a)&0.63\\\hline
\end{array}
$$

\caption{Same as Table \ref{co4} for Co inside the (8,8) tube.}\label{co8in}
\end{table*}

\begin{table*}[ptb]
 $$
\begin{array}{|c|c|c|c|c|c|c|c|c|c|c|}\hline
orbital&\epsilon_{\uparrow}&\epsilon_{\downarrow}&\delta_{\uparrow}&\delta_{\downarrow}&\Gamma_{PDOS}&\Gamma&\phi&\delta_{NRG}&g_{NRG}&g_{DFT}\\\hline
d_{z^2}(es)&-3.08&0.23&-1.443&0.363&0.066&0.060&0.149&-0.09&&\\
s(es)&0.01&1.64&&&0.039&(0.039)&&&&\\
d_{xz}(os) &-2.10&1.10&0.188&0.308&&0.092&0.230&-1.57&0.03(s)&0.50\\
d_{xy}(ea) &-2.90&0.70&0.197&0.333&0.065&0.082&0.224&-1.29&&\\
d_{yz}(oa) &-2.88&-1.24&0.100&-0.039&&0.301&0.202&&0.09(a)&0.93\\\hline
\end{array}
$$

\caption{Same as Table \ref{co4} for the Fe/(4,4) system.}\label{fe4}
\end{table*}

\begin{table*}[ptb]
 $$
\begin{array}{|c|c|c|c|c|c|c|c|c|c|c|}\hline
orbital&\epsilon_{\uparrow}&\epsilon_{\downarrow}&\delta_{\uparrow}&\delta_{\downarrow}&\Gamma_{PDOS}&\Gamma&\phi&\delta_{NRG}&g_{NRG}&g_{DFT}\\\hline
d_{z^2}(es)&-1.96&-0.06&0.112&-0.681&0.059&0.057&0.141&-0.09&&\\
d_{xz}(os)&-1.60&1.14&0.135&0.225&&0.062	&0.172&1.57&0.01(s)&0.69\\
d_{xy}(ea)&-1.72&1.12&0.137&0.200&&0.044&0.162&1.57&&\\
d_{yz}(oa)&-2.08&-0.61&0.092&-0.027&&0.103&0.141&&0.00(a)&0.98\\\hline
\end{array}
$$

\caption{Same as Table \ref{co4} for the Fe/(8,8) system.}\label{fe8}
\end{table*}

\begin{table*}[ptb]
 $$
\begin{array}{|c|c|c|c|c|c|c|c|c|c|c|}\hline
orbital&\epsilon_{\uparrow}&\epsilon_{\downarrow}&\delta_{\uparrow}&\delta_{\downarrow}&\Gamma_{PDOS}&\Gamma&\phi&\delta_{NRG}&g_{NRG}&g_{DFT}\\\hline
d_{z^2}(es)&-1.33&-0.27&-0.197&-0.905&&0.363&0.073&-0.38&&\\
d_{xz}(os) &-1.35&1.06&0.007&0.215&&0.126&0.100&-1.57&0.16(s)&0.56\\
d_{xy}(ea) &-0.84&0.90&-0.241&0.588&0.442&0.396&0.222&1.54&&\\
d_{yz}(oa) &-1.69&-0.47&0.083&-0.115&&0.129&0.158&&0.00(a)&0.76\\\hline
\end{array}
$$

\caption{Same as Table \ref{co4} for Fe inside the (4,4) tube.}\label{fe4in}
\end{table*}

\begin{table*}[pt]
 $$
\begin{array}{|c|c|c|c|c|c|c|c|c|c|c|}\hline
orbital&\epsilon_{\uparrow}&\epsilon_{\downarrow}&\delta_{\uparrow}&\delta_{\downarrow}&\Gamma_{PDOS}&\Gamma&\phi&\delta_{NRG}&g_{NRG}&g_{DFT}\\\hline
d_{z^2}(es)&-1.62&-0.32&0.008&-0.318&&0.112&0.086&-0.13&&\\
d_{xz}(os)&-1.59&1.13&0.073&0.193&&0.081	&0.124&-1.54&0.04(s)&0.88\\
d_{xy}(ea)&-1.56&1.12&0.039&0.240&&0.134&0.125&-1.51&&\\
d_{yz}(oa)&-1.70&-0.27&0.059&-0.214&&0.009&0.111&&0.01(a)&0.90\\\hline
\end{array}
$$

\caption{Same as Table \ref{co4} for Fe inside the (8,8) tube.}\label{fe8in}
\end{table*}

\end{document}